\documentclass{nature}
\usepackage[top=0.8in, bottom=1in, left=0.8in, right=0.8in]{geometry}
\usepackage{rotating}
\usepackage{caption}
\usepackage{graphicx}
\usepackage{amsmath}
\usepackage{amssymb}
\usepackage{xcolor}
\usepackage{threeparttablex, tablefootnote}
\usepackage{longtable}
\usepackage{booktabs}
\usepackage{lineno}

\title{Strong spectral features from asymptotic giant branch stars in distant quiescent galaxies}

\author{\large Shiying Lu$^{1,2,3}$,
Emanuele Daddi$^{2}$,
Claudia Maraston$^{4}$, 
Mark Dickinson$^{5}$, 
Pablo Arrabal Haro$^{5}$,
Raphael Gobat$^{6}$,
Alvio Renzini$^{7}$,
Mauro Giavalisco$^{8}$, 
Micaela B. Bagley$^{9}$,      
Antonello Calabr{\`o}$^{10}$,  
Yingjie Cheng$^{8}$,
Alexander de la Vega$^{11}$,
Chiara D'Eugenio$^{12,14}$,       
David Elbaz$^{2}$,
Steven L. Finkelstein$^{9}$,                      
Carlos G{\'o}mez-Guijarro$^{2}$, 
Qiusheng Gu$^{1,3}$,
Nimish P. Hathi$^{13}$,
Marc Huertas-Company$^{12,14,15}$,
Jeyhan S. Kartaltepe$^{16}$,     
Anton M. Koekemoer$^{13}$,
Aur{\'e}lien Henry$^{17}$,
Yipeng Lyu$^{2}$,
Benjamin Magnelli$^{2}$,
Bahram Mobasher$^{11}$,        
Casey Papovich$^{18,19}$,    
Nor Pirzkal$^{20}$,  
R. Michael Rich$^{21}$,
Sandro Tacchella$^{22,23}$
and L. Y. Aaron Yung$^{13}$
\vspace{8pt}}

\begin{document}

\maketitle

\begin{affiliations}
\small
\item School of Astronomy and Space Science, Nanjing University, Nanjing, 210093, China.
\item Universit{\'e} Paris-Saclay, Universit{\'e} Paris Cit{\'e}, CEA, CNRS, AIM, Paris, 91191,France
\item Key Laboratory of Modern Astronomy and Astrophysics (Nanjing University), Ministry of Education, Nanjing, 210093, China
\item Institute of Cosmology and Gravitation, University of Portsmouth, Dennis Sciama Building, Burnaby Road, Portsmouth PO1 3FX, UK
\item NSF's National Optical-Infrared Astronomy Research Laboratory, 950 N. Cherry Ave. Tucson, AZ 85719, USA
\item Instituto de F{\'i}sica, Pontificia Universidad Cat{\'o}lica de Valpara{\'i}so, Casilla 4059, Valpara{\'i}so, Chile
\item INAF-Osservatorio Astronomico di Padova, Vicolo dell'Osservatorio 5, I-35122 Padova, Italy
\item University of Massachusetts Amherst, 710 North Pleasant Street, Amherst, MA 01003-9305, USA
\item Department of Astronomy, The University of Texas at Austin, Austin, TX, USA
\item INAF-Osservatorio Astronomico di Roma, via di Frascati 33, 00078 Monte Porzio Catone, Italy
\item Department of Physics and Astronomy, University of California, 900 University Ave, Riverside, CA 92521, USA
\item Instituto de Astrofísica de Canarias (IAC), 38205 La Laguna, Tenerife, Spain
\item Space Telescope Science Institute, 3700 San Martin Dr., Baltimore, MD 21218, USA
\item Universidad de la Laguna, La Laguna, Tenerife, Spain
\item Universit\'e Paris-Cit\'e, LERMA - Observatoire de Paris, PSL, Paris, France
\item Laboratory for Multiwavelength Astrophysics, School of Physics and Astronomy, Rochester Institute of Technology, 84 Lomb Memorial Drive, Rochester, NY 14623, USA
\item Department of Physics, University of California Merced, 5200 Lake Road, Merced, CA 95343, USA
\item Department of Physics and Astronomy, Texas A\&M University, College Station, TX, 77843-4242, USA
\item George P.\ and Cynthia Woods Mitchell Institute for Fundamental Physics and Astronomy, Texas A\&M University, College Station, TX, 77843-4242, USA
\item ESA/AURA Space Telescope Science Institute
\item Department of Physics and Astronomy, University of California, Los Angeles, CA, 90095, USA
\item Kavli Institute for Cosmology, University of Cambridge, Madingley Road, Cambridge, CB3 0HA, UK
\item Cavendish Laboratory, University of Cambridge, 19 JJ Thomson Avenue, Cambridge, CB3 0HE, UK
\end{affiliations}

\begin{abstract}
Dating the ages and weighting the stellar populations in galaxies are essential steps when studying galaxy formation through cosmic times. Evolutionary population synthesis models with different input physics are used for this purpose. Moreover, the contribution from the thermally pulsing asymptotic giant branch (TP-AGB) stellar phase, which peaks for intermediate-age 0.6-2 Gyr, has been debated for decades. Here we report the detection of strong cool-star signatures in the rest-frame near-infrared spectra of three young ($\sim$1Gyr), massive ($\sim 10^{10}M_\odot$) quiescent galaxies at large look-back time, z=1-2, using JWST/NIRSpec. The coexistence of oxygen- and carbon-type absorption features, spectral edges and features from rare species, such as vanadium and possibly zirconium, reveal a strong contribution from TP-AGB stars. Population synthesis models with a significant TP-AGB contribution reproduce the observations better than those with a weak TP-AGB, which are commonly used. These findings call for revisions of published stellar population fitting results, as they point to populations with lower masses and younger ages and have further implications for cosmic dust production and chemical enrichment. New generations of improved models are needed, informed by these and future observations.
\end{abstract}

\section{Introduction} 
The thermally pulsing asymptotic giant branch (TP-AGB) -- an advanced evolutionary stage of low- and intermediate-mass stars (0.8-10$M_\odot$) -- is a potentially essential component of stellar population models. Quantifying the TP-AGB contribution to the light of stellar populations has a long history. The first estimate goes back over 30 years\cite{Renzini+1981}, which was based on the \textit{fuel consumption theorem}\cite{Renzini+Voli+1981}, predicted that the TP-AGB contribution to the bolometric light of a simple stellar population (SSP) could approach or even exceed $\sim$50\% at intermediate ages. Subsequent models estimated a contribution of up to $\sim$80\% of the integrated near-infrared (NIR) emission of stellar populations with ages in the range of 0.2 $\lesssim$$t$$\lesssim$ 2 Gyr\cite{Maraston+1998, Maraston+05, Salaris+14, Maraston+06, Tonini+09}. 

The actual energetic contribution and spectral composition of this phase, however, remain highly controversial, with different models adopting distinctively different prescriptions for both the energetics and the stellar spectra\cite{Bruzual+Charlot+03, Maraston+05, Marigo+08, Conroy+09}. These uncertainties stem from two complications typical of the TP-AGB phase, namely its double-shell burning regime leading to instabilities and `thermal pulses' on short timescales alongside a strong mass-loss, both requiring calibration of free parameters such as convection, mixing and mass loss, with observations\cite{Noel+13, Pastorelli+2020}. As a consequence, stellar evolutionary computations, which are the backbone of evolutionary population synthesis models, typically terminate at the end of the early AGB  phase, and the inclusion of the TP-AGB in integrated models is performed separately either using specific TP-AGB tracks\cite{Marigo+08, Salaris+14} or fuel-consumption-theorem estimated TP-AGB energetics\cite{Maraston+05}. 

Historically, the early Maraston (M05\cite{Maraston+05}) models based on the \textit{fuel consumption theorem}\cite{Renzini+Voli+1981} pioneered the incorporation of the TP-AGB phase into population synthesis models by calibrating the energetics of TP-AGB tracks with observed photometric data and adopting observed oxygen-rich (O-rich) and carbon-rich (C-rich) spectra\cite{Lancon+2000, Lancon+02}. These M05 models, which forecast the onset of the TP-AGB phase as a sharp transition (e.g., a rapid increase in the TP-AGB bolometric and NIR contribution to the energetics of stellar populations\cite{Maraston+1998}) around 0.2 Gyr and lasting up to 2 Gyr, are referred to as ``TP-AGB-heavy''. A similar approach is found in the Conroy (C09\cite{Conroy+09}) models, but with two modifications, namely that there is a substantial reduction of the energetics and a shift towards older ages (1-3 Gyr) for the relevant population epoch, which dilute the TP-AGB `phase transition'. The C09 models are referred to as ``TP-AGB-light''. Following the availability of new data for the calibration and adopting some of the C09 arguments, the calibration of the TP-AGB contribution as a function of age was revised using data for globular clusters in the Magellanic Clouds\cite{Noel+13}. These updated versions of the M05 models are the M13 models, which are referred to as ``TP-AGB-mild''. They have a reduced TP-AGB fuel consumption and a slightly older onset age (600 Myr) with respect to M05 but still predict a sharp transition for the TP-AGB onset and sizable fuel consumption. Finally, the widely used Bruzual \& Charlot (BC03\cite{Bruzual+Charlot+03}) models, which are based on \textit{isochrone synthesis} of stellar evolutionary tracks\cite{Girardi+00}, adopt different energetics and spectra for the TP-AGB phase. As this results in a negligible TP-AGB contribution to the integrated spectrum, the BC03 models are referred to as ``TP-AGB-poor''. 

Many studies have attempted to observationally constrain the contribution of the TP-AGB phase by fitting spectra and photometric spectral energy distributions (SEDs) of different types of galaxies (post-starbursts, Seyferts, spirals, and passive) over a range of redshifts, with conflicting results\cite{Kriek+10, Zibetti+13, Riffel+15, Capozzi+16}. Quiescent high-redshift galaxies (mass-weighted ages of $\sim$1 Gyr produce the peak contribution of TP-AGB-stars to the NIR) are ideal laboratories but until now, imaging data have been available only at high redshift. High-quality NIR galaxy spectra dominated by TP-AGB stars are still missing: ground-based spectra below 2.3$\mu$m are impacted by atmospheric emission and absorption, and NIR grism spectra from Hubble Space Telescope (HST) lack sufficient sensitivity. The James Webb Space Telescope (JWST) has the wavelength coverage and sensitivity to finally settle this issue.

\section{Results}
\subsection{Selection of quiescent galaxies\\} 
We searched for distant quiescent galaxies serendipitously included (as fillers) in the Cosmic Evolution Early Release Science (CEERS) program of the near-infrared spectrograph (NIRSpec) onboard JWST\cite{Finkelstein+23, Arrabal+Haro+23} and during a subsequent NIRSpec follow-up program (DD-2750)\cite{Arrabal+Haro+23_natr}. We started identifying 1084 quiescent galaxy candidates at $z_{phot}$$>$1 in the  CANDELS extended Groth strip\cite{Grogin+11, Koekemoer+11} catalog\cite{Stefanon+17}, based on the UVJ diagram\cite{Whitaker+11} and SED modelling (adopting exponentially declining star-formation history, namely SFH$\propto$$e^{-t/\tau}$ and requiring $t/\tau$$>$1.0), in which $\tau$ is the e-folding quenching timescale and the $t$ is the time since the onset of star formation. The $z$$>$1 limit ensures that the age-sensitive spectral region around rest-frame 4000\AA\ break is covered by JWST/NIRSpec with good sensitivity. We cross-matched these photometric candidates with the 1491 galaxies already with secure NIRSpec spectroscopic redshift. This resulted in 31 galaxies whose spectra we visually inspected. Some spectra have low signal-to-noise ratios (SNR) because the galaxies are faint or poorly centered in the NIRSpec's micro-shutter apertures. Other spectra had strong emission lines consistent with active star formation or an active galactic nucleus. A few have redshifts ($z$$>$3) that are too high to provide adequate coverage of the relevant rest-frame NIR wavelengths. We retained three galaxies, IDs 8595, 9025, and D36123, which have redshifts of 1$<$$z$$<$2, and whose spectra clearly show the D4000 break and no obvious emission lines and have relatively high SNR.

\subsection{The primary spectrum of galaxy D36123\\} 
\begin{figure*}
\begin{center}
\includegraphics[width=0.9\linewidth]{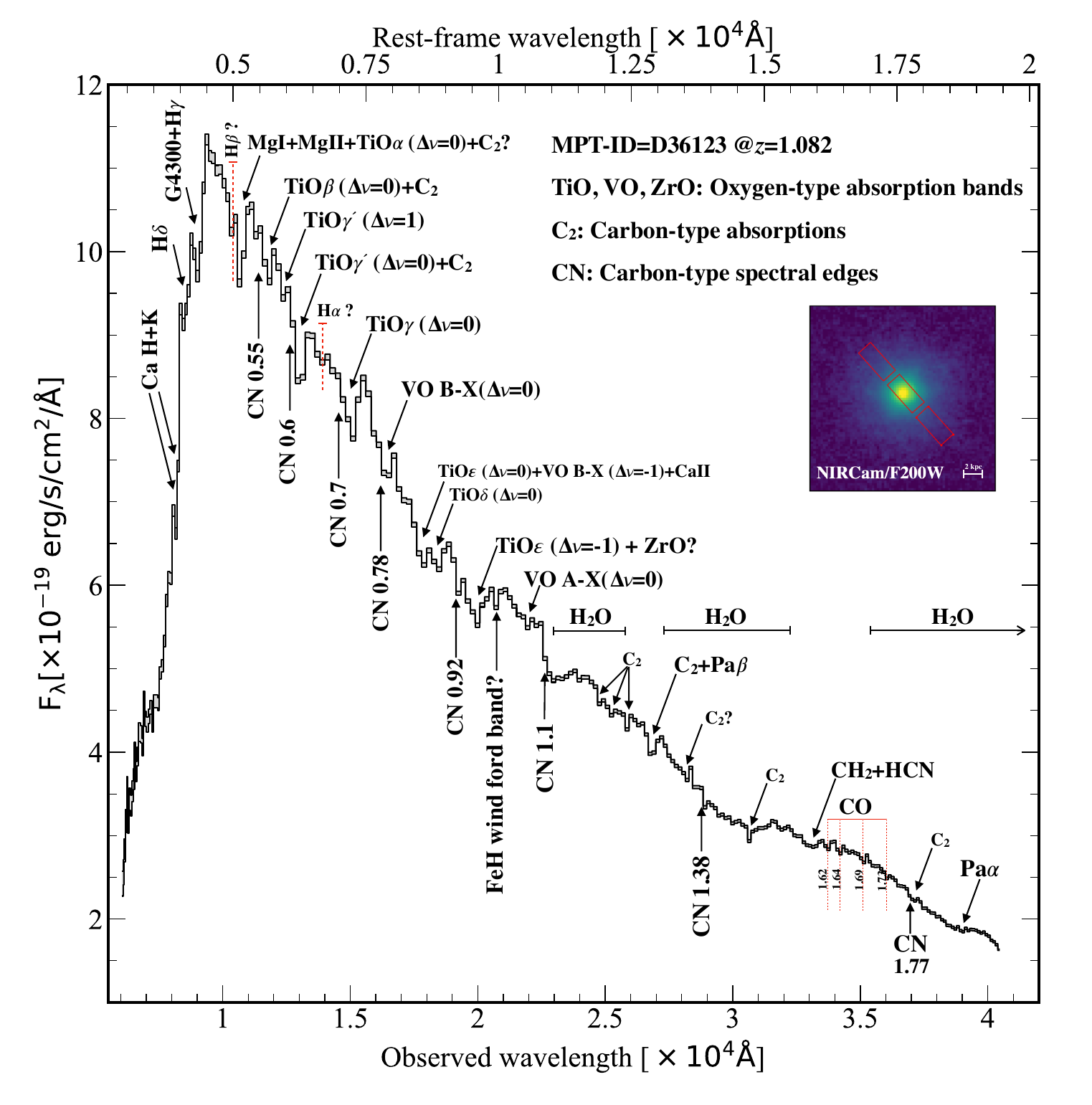}
\end{center}
\vspace{-0.3truecm}  
\caption{\small \textbf{ NIR rest-frame spectrum of the quiescent galaxy D36123 at $z$=1.082.}  The upper and lower histograms (filled in gray) display the 1$\sigma$ uncertainty range (notice the high SNR). The rest-frame wavelengths are calculated from the redshift corresponding to the best fit of M13 models. Black arrows point to identifications of oxygen-type and carbon-type absorption and the detections of deep CN edges (see also  Supplementary Table~\ref{SI_tab1}). Both sets of features are characteristic of carbon-rich and oxygen-rich stars in the TP-AGB phase. The question mark indicates uncertain features. The NIRCam image in the F200W band is shown in the inset together with the NIRSpec shutters layout (each shutter is $0.2''\times 0.46''$).}\label{fig01}
\end{figure*}

Galaxy D36123 exhibits an exceptionally high-quality PRISM spectrum with an average of SNR$\gtrsim$187 per spectral pixel (Fig.\ref{fig01}; plotted as a double histogram showing the $\pm1\sigma$ noise). The relatively bright flux (mag$_{\rm AB}$=21.4 at 3.6$\mu$m), steep radial surface brightness profile (S{\'e}rsic $n$$\sim$4, see Extended Data Fig.\ref{ext_fig01}), longer exposure time and accurate centering into the micro-shutters, which minimizes the effect of path loss and aperture correction (Methods), are the reasons for the notable quality of this spectrum.

\begin{figure*}[hbtp]
\begin{center}
 \includegraphics[width=\linewidth]{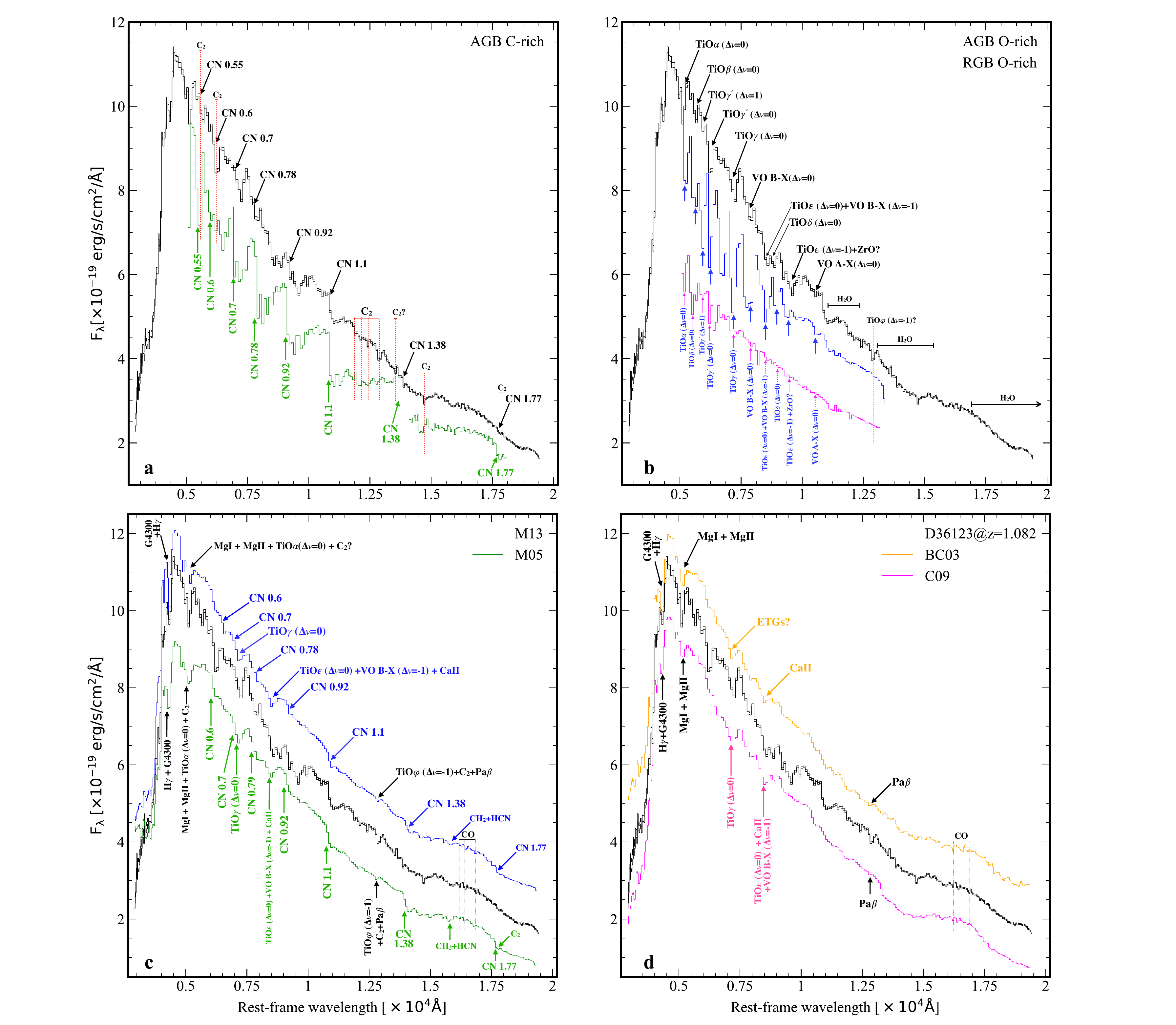}
 \end{center}
 \vspace{-0.3truecm}  
 \caption{\small \textbf{Identificated features and best-fitting models for D36123.} {\bf a,b,} Identified TP-AGB features (solid arrows) based on Milky-Way stars. {\bf a,}The deep CN edges and carbon-containing molecules are confirmed by comparing them to a C-rich (C1) star\cite{Lancon+02} (green) in the TP-AGB phase. {\bf b,} The TiO, VO, and ZrO absorptions were identified as stemming from an O-rich star in the coldest bin\cite{Lancon+2000} (blue) in the TP-AGB phase. The O-rich static giants\cite{Verro+22} with (I-K)$_{\rm color}$=2.01 in the RGB phase are plotted in magenta, with dashed arrows indicating the positions of unmatched features. For clarity, the stellar spectra are stretched to follow the continuum shape of our high redshift target. \textbf{c,d,} Fitting the D36123 spectrum with stellar population models. {\bf c,} The TP-AGB-mild (M13, blue) and TP-AGB-heavy (M05, green) models. {\bf d,} The TP-AGB-poor (BC03, orange) and TP-AGB-light (C09, magenta) models. All best-fitting models are additively shifted for clarity. Black arrows/labels highlight features identified by all four models at the same absorption positions. Two absorption features in orange (magenta) are identified by BC03 (C09) model. Other CN edges and oxygen-type absorptions can be identified by M13 and M05 models. The black histograms in all panels show the rest-frame spectrum of D36123 and its errors, adopting the redshift $z$=1.082 from M13 best fit. }\label{fig02}
\vspace{-12pt}
\end{figure*}

\begin{figure*}[htbp]
 \begin{center}
 \includegraphics[width=0.9\linewidth]{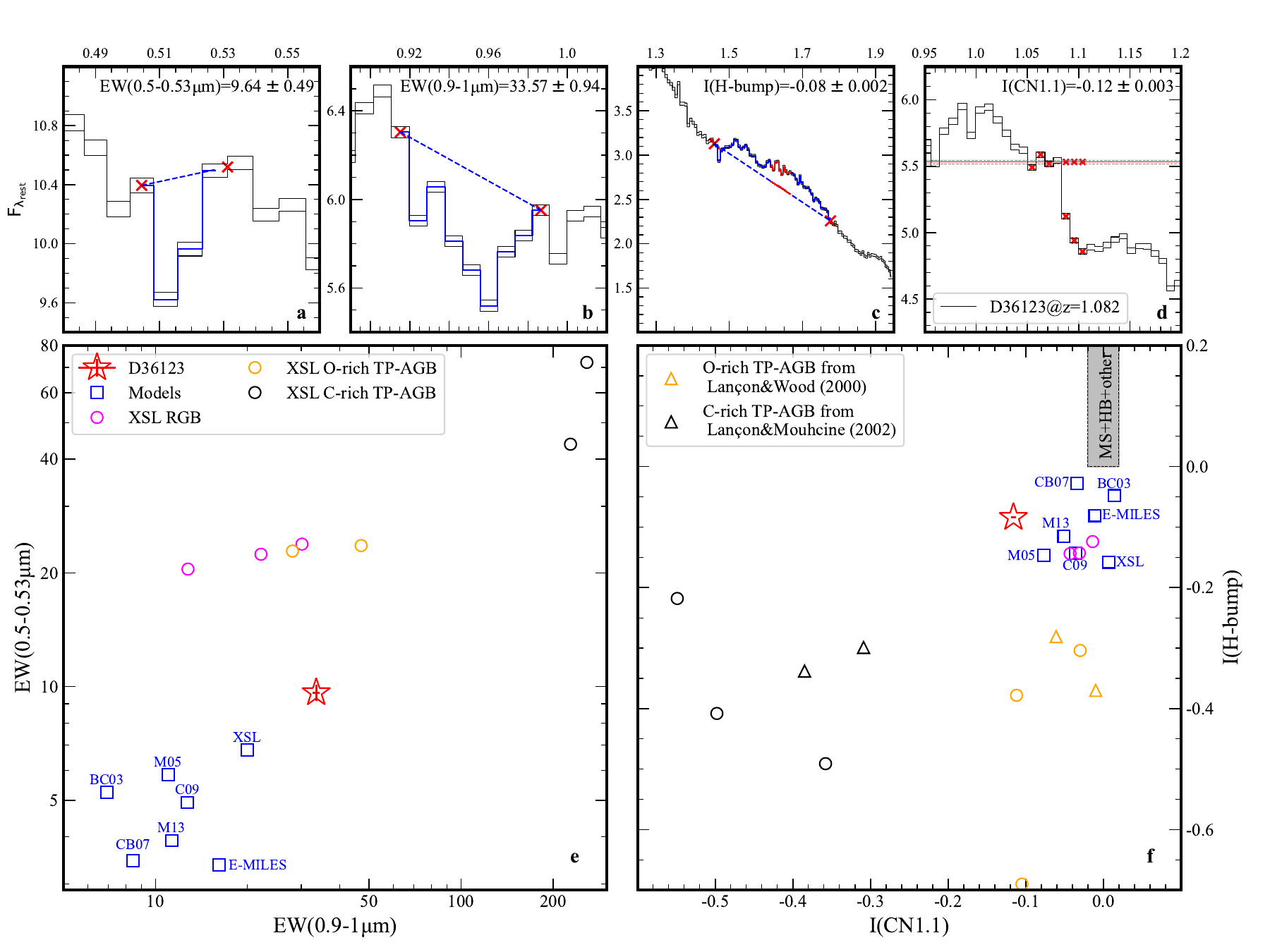}
 \end{center}
 \vspace{-0.3truecm}  
\caption{\small \textbf{Specific features and comparison with models and stellar templates.} \textbf{a-d,} Specific features of galaxy D36123 in the rest frame. EW at 0.5-0.53$\mu$m ({\bf a}) and 0.9-1$\mu$m ({\bf b}). Strengths of the H-band bump at $\sim$1.6$\mu$m ({\bf c}) and of the CN1.1 edge at $\sim$1.1$\mu$m ({\bf d}). \textbf{e,f,} Comparisons of the best-fitting models and cool stars. {\bf e,} Relationship between EW(0.5-0.53$\mu m$) and EW(0.9-1.0$\mu$m). {\bf f,} Relation between two indexes, I(H-bump) and I(CN1.1). Values for the indices for models and cool stars are listed in Extended Data Table 1. Cool stars include the O-rich static giants (i.e., RGB) and C-/O-rich TP-AGB stars from the averaged spectra of XSL\cite{Verro+22}, shown by circles in panel {\bf e}. Other C-/O-rich TP-AGB stars from other libraries\cite{Lancon+2000, Lancon+02} are shown by triangles in panel {\bf f}. The grey shaded area in {\bf f} shows the distribution of the main sequence (MS), horizontal branch (HB), and others during the stellar phases.}\label{fig03}
\vspace{-12pt}
\end{figure*}

A large number of features are clearly detected at rest-frame wavelengths from 0.3 to 2.0 $\mu$m, for which we propose identifications (see labels in Fig.\ref{fig01} and Supplementary Table~\ref{SI_tab1}). Besides common absorption features at $\lambda_{\rm rest}$$<$5500\AA\ typical of quiescent galaxies, such as Ca H+K, H$\delta$, H$\gamma$+G4300 and absorption from heavy elements such as Mg, Ca, and Fe, which have been observed in quiescent galaxies at high redshift\cite{Onodera+15}, the spectrum exhibits numerous absorption features and band edges. The relatively low spectral resolution (R$\approx$30-300) of the NIRSpec PRISM helps in the detection of these broad features. The proposed identifications were obtained by comparison with empirical libraries of C-rich\cite{Lancon+02} and O-rich\cite{Lancon+2000} TP-AGB stars (see Fig.\ref{fig02}a,b). These features are related to diverse stellar types in the AGB stellar phases, that is early AGB and TP-AGB\cite{Lancon+01, Habing+04, Riffel+07, Rayner+09}. The deep CN edges in Fig.\ref{fig02}a (most striking is the CN1.1$\mu$m edge in Fig.\ref{fig01}, where the continuum suddenly drops down by $\sim$15\%) are particularly enhanced in N-type carbon stars\cite{Lancon+01, Habing+04, Riffel+07}, which are produced by the third dredge-up. These deep CN edges, therefore, demonstrate the presence of the contribution by TP-AGB stars (see CN1.1 index in Fig.\ref{fig03}d,f). On the other hand, the TiO and VO absorption bands (Fig.\ref{fig02}b) are produced in cool M-type stars in either the red giant branch (RGB) or AGB\cite{Habing+04}, but the equivalent width (EW) measured for specific features at 0.5-0.53 and 0.9-1$\mu$m and the strength of the 1.6$\mu$m H bump are all larger than what would be expected for the RGB alone, when diluted by the rest of the stellar phases, which are featureless at these wavelengths (Fig.\ref{fig03}e,f). 
Multiple TiO absorptions concomitant with CN and C$_2$ features have been observed in a galaxy's spectrum, pointing to the simultaneous presence of O-rich and C-rich cool stars. This was naturally expected for a broad distribution of stellar metallicities, which must exist in massive galaxies, such that the metal-poor stars evolve preferentially into C-rich TP-AGB stars, and the metal-rich ones into O-rich (M Type) TP-AGB stars\cite{Renzini+Voli+1981}. As V and Zr are produced by only \textit{s}-process nucleosynthesis, those VO and ZrO features (Fig.\ref{fig02}b), indicative of a recent third dredge-up episode, are expected only in the coolest TP-AGB stars\cite{Habing+04, Rayner+09}, demonstrating the prominence of the TP-AGB phase in the light emitted by this $z$$\sim$1 quiescent galaxy.

We subsequently fitted the spectrum of D36123, employing four models -- BC03, C09, M13, and M05. These span the full range from TP-AGB poor to heavy (see Fig.\ref{fig02}c,d and Table~\ref{tab1}). The first two are the standard for deriving the physical properties of high-redshift galaxies. We used delayed exponentially declining SFH$\propto$$(t/\tau^2)e^{-t/\tau}$ with a large variety of timescales and stellar ages, interpolated over a grid of metallicities, and allowed for dust extinction. When convolving the models to match the resolution of observed spectra, we considered the instrumental resolution, the wavelength sampling of the spectral template models and the estimated stellar velocity dispersion of the galaxy (Methods). We found that our model fitting results were robust against (1) more complex SFHs allowing for bursts and sudden truncations, (2) a non-parametric SFH obtained as linear combinations of models with different ages and metallicities, (3) alternative attenuation curves or without considering the reddening by dust, (4) variations to the aperture correction recipe, (5) modifications of the spectral extraction apertures over 3 or 5 pixels, (6) modelling of emission lines. The minimum reduced $\chi_{\rm R}^2$ from each model was barely affected when these variations are considered. 

The TP-AGB-mild and -heavy models (M13 and M05) can reproduce the largest fraction of the detected features, whereas the TP-AGB-poor and -light models predict much weaker spectral features in the NIR, as Fig.\ref{fig02}c,d and Fig.\ref{fig03}e,f illustrate (also Extended Data Table~\ref{ext_tab1}). The TP-AGB-mild model, M13, provided the best overall fit and also identified the majority of TP-AGB features (four TiO absorptions and most CN edges). However, even in the M13 best-fit spectrum, some strong spectral features are clearly not present (or much weaker than in the observed spectrum), like several TiO and C$_2$ absorptions. The strong CN1.1$\mu$m spectral edge is best reproduced by the M13/M05 models (Fig.\ref{fig03}f). Interestingly, the MgII band at 0.52$\mu$m, one of Lick  indices\cite{Faber+1985}, is much weaker in all models than in the data (Fig.\ref{fig03}e), pointing to a strong contribution from TiO bands all the way to 0.5$\mu$m, as expected for M-type stars (and, possibly, to [$\alpha$/Fe]-enhancement\cite{Thomas+05}). Overall, M05 produced the worst fit due to clear discrepancy at $\lambda_{\rm rest}$$<$5000\AA\ (see Extended Data Fig.\ref{ext_fig02}), but still had significantly more matching TP-AGB features in the NIR than the C09 and BC03 models (Fig.\ref{fig02}c,d), noticeably the long-sought C$_2$ absorption around 1.75$\mu$m. Moreover, when fitting the spectrum over the restricted range $\lambda_{\rm rest}$$<$5000\AA\, all models converged to a mass-weighted age of about 1~Gyr, which is, indeed, the age when TP-AGB features are expected to nearly peak\cite{Maraston+05, Noel+13}. Therefore, we conclude that the treatment of the TP-AGB in the models most affects the goodness of fit in our $z$$\approx$1 quiescent galaxies. Even for the top performing model, M13, the minimum reduced $\chi_{\rm R}^2$$\approx$39 was large, indicating that this model also lacks stellar ingredients that are present in the galaxy and are necessary.  

As mentioned, some observed features were also found in the coolest phase of (M-type) RGB stars (Fig.\ref{fig02}b) and the red supergiant phase of massive stars, which have ages around 10 Myr\cite{Lancon+2000, Lancon+01, Rayner+09}. We found that red supergiant stars did not contribute substantially due to the inferred mass-weighted age ($\sim$1 Gyr) and negligible residual star-formation rate (SFR) when we fitted only the optical spectrum (see Table~\ref{tab1}), implying a negligible contribution from short-lived massive stars. M-type RGB stars, on the other hand, certainly contribute to the NIR spectrum at these ages, to a degree that is, however, model-dependent because the onset and development of the RGB phase depends on the overshooting assumed in stellar tracks\cite{Ferraro+04, Maraston+05}. The full RGB contribution to the total light in the K-band around 1 Gyr is $\sim$20-30\% according to M05 models and $\sim$10-20\% for BC03-type models.
The upper part of the RGB where M-giants form is a further fraction of this. Therefore, although they contribute, M-type RGB stars are not expected to dominate the spectrum. This expectation is confirmed quantitatively by the measured indices in Fig.\ref{fig03}f, whose strengths are comparable or even larger than those for individual RGB stars, which are necessarily diluted by the composite populations in galaxies. Therefore, many TP-AGB stars with strong features are needed to match the observed strengths of the features in the galaxy.

The best-fitting physical parameters presented in Table~\ref{tab1} exemplify the impact of the chosen model on the derived galaxy properties. For instance, comparing results between BC03 and M13 in the full rest-frame spectral range, there is only a 30\% discrepancy in stellar mass, but the impact on ages is strong, with stellar age difference $\Delta t$$\approx$1 Gyr (the M13 age is almost three times lower than the BC03 age), despite the higher metallicity preferred by the BC03 fit. These age differences affect the derived formation epochs and evolutionary timescales of galaxies. 

We also fitted some other models. Two were entirely based on empirical stellar libraries (X-shooter Spectral Library (XSL)\cite{Verro+22} and E-MILES\cite{Vazdekis+15}; see Extended Data Fig.\ref{fig03}a,b and Extended Data Table~\ref{ext_tab2}), but there were no significant improvements in the quality of fit or reproduction of features (Fig.\ref{fig03}e,f). These models returned old ages $>$1.5Gyr, with a subsolar metallicity best fit for E-MILES. The EW indices for XSL are closer to what was observed for D36123, but the model performed less well on CN1.1$\mu$m. The CB07 models (an updated version of BC03; Methods) behaved like M05 in providing a worse global fit but better match to specific deep features (for example, TiO$\gamma$ at 0.73$\mu$m, Extended Data Fig.\ref{ext_fig03}c).

\subsection{The spectra of galaxies 8595 and 9025 \\} 
\begin{figure*}[hbtp]
\begin{center}
\includegraphics[width=0.9\linewidth]{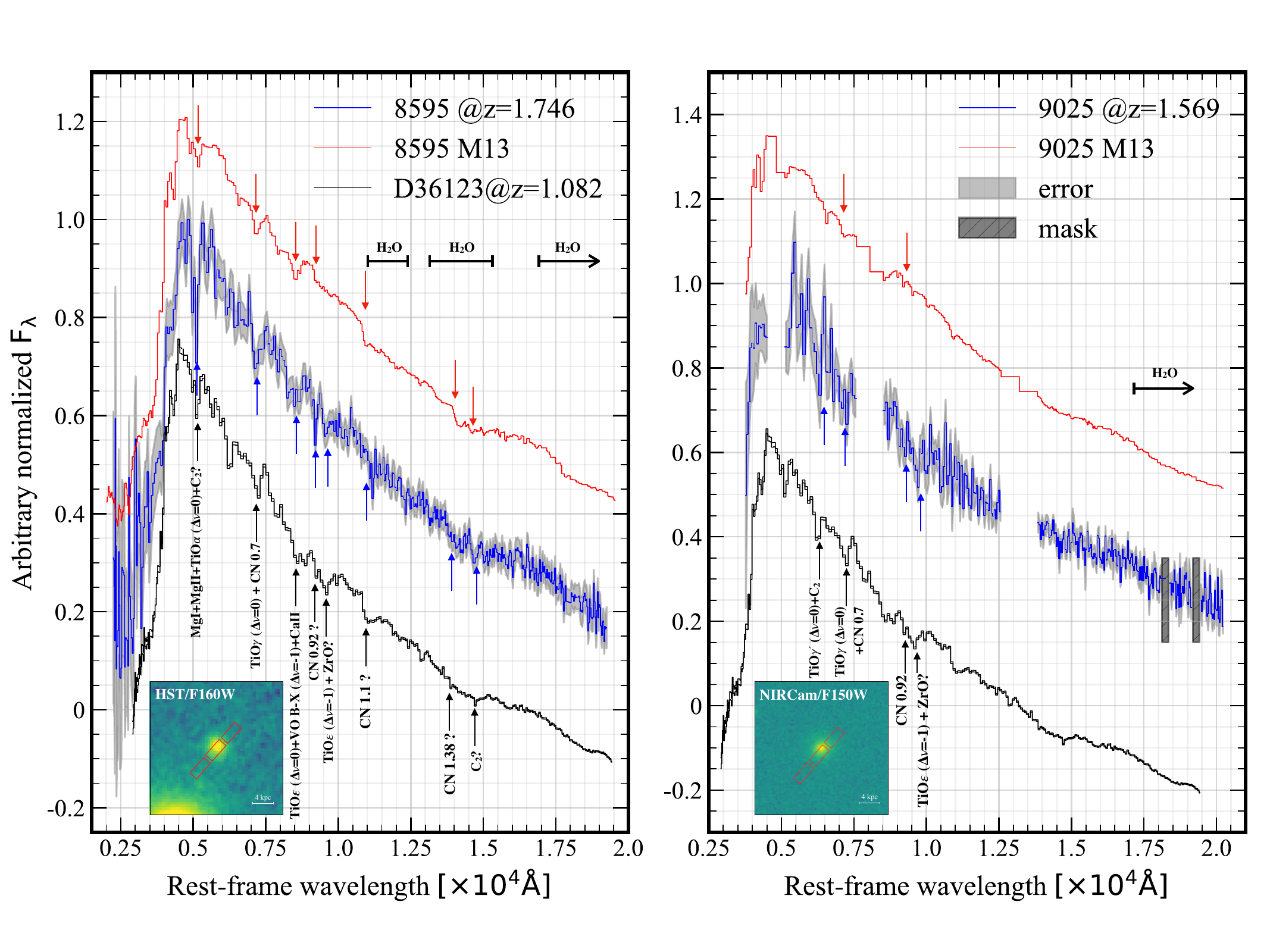}
\end{center}
\vspace{-0.3truecm}  
\caption{\small \textbf{NIR spectra and identified features for the other two galaxies.} The left and right panels are for galaxies 8595 and 9025, respectively. The observed spectrum (blue) with error (grey-shaded region) is compared with the M13 best-fitting model (red) and the spectrum of the bright galaxy D36123 (black) to identify features. The comparison spectra are shifted up or down additively for clarity. Blue arrows indicate matched features (labelled) in the spectra. Most features predicted by the M13 model are indicated by red arrows. Tentative identifications are shown by a question mark. Inserts, HST/F160W (NIRCam/F150W) image for galaxy 8595 (9025) with NIRSpec shutters overlaid. Note that the observed spectrum of galaxy 9025 with the median resolution has been resampled to the PRISM spectral pixel grid to increase SNR and for clarity. Some contaminated regions are blanked in the redder part of the spectrum of galaxy 9025. Detector defects in the spectrum of object 9025 are marked (grey rectangles; $\lambda_{\rm rest} \approx$ 1.81-1.84$\mu$m and 1.92-1.94$\mu$m).}\label{fig04}
\vspace{-12pt}
\end{figure*}

The spectra of the other two fainter quiescent galaxies, 8595 (mag$_{\rm AB}$=24 at 3.6$\mu$m, observed with the PRISM) and 9025 (mag$_{\rm AB}$=23 at 3.6$\mu$m, observed with the medium-resolution gratings but rebinned to the PRISM resolution), have substantially lower SNR, $\sim$15 and 7. Nevertheless, the SNR is still sufficient to detect  
individual absorption features in object 8595 (Fig.\ref{fig04}, left). Some of the strongest features were observed for object 9025 (Fig.\ref{fig04}, right), including TiO absorption lines, CN spectral edges and, tentatively, a C$_2$ absorption feature.

Table~\ref{tab2} presents the results of the spectral fitting of objects 8595 and 9025. For object 8595, we carefully accounted for a neighbouring bright galaxy when measuring its photometry to estimate spectral aperture corrections, following a similar approach to what was done for object D36123 (Methods). Again, M13 yielded the best-fitting spectral fits for both galaxies, with $\chi^2_{\rm R}$$\approx$0.9 for object 8595, and $\chi^2_{\rm R}$$\approx$1.7 for object 9025. The derived stellar masses differ at most by $0.16$ dex between the M13 and BC03 models, but the best-fitting ages were even more strongly dependent on the specific model used ($\Delta t$$\approx$0.1-1 Gyr). In Table~\ref{tab2}, the values $\Delta \chi^2$ denote the $\chi^2$ difference with respect to the M13 model. The small cumulative probability $P(\Delta \chi^2)$ suggests the preference for the M13 model is statistically significant in the overall spectral fitting for these two galaxies.

\section{Discussion}  
Overall, we conclude that the spectra of distant quiescent galaxies studied here show clear evidence of substantial contribution of both C- and M-type TP-AGB stars. The models with significant TP-AGB contribution, such as M13 and M05, are in better agreement with NIR features in the galaxies' spectra, whereas TP-AGB-mild models (M13) can fit their full spectra better, supporting our conclusion that the TP-AGB is the primary model ingredient driving the quality of the fits.

Previous work\cite{Riffel+15} based on NIR spectra of local spiral and Seyfert galaxies reached similar conclusions, favouring M05 models over BC03 models because of the TP-AGB component. Those local spectra cover a wide wavelength range like our high-redshift data and display various features, several of which can be attributed only to TP-AGB stars. Previously, detection of the CN 1.1$\mu$m band in the NIR spectra of local active galactic nuclei has been interpreted as a signature of intermediate age populations evolving along the TP-AGB phase\cite{Riffel+07}. Although these local observations are notable, JWST spectroscopy can observe massive galaxies at high redshift. Their low spread in stellar generations allows for their stars to evolve through the TP-AGB phase in sync at age $\sim$1 Gyr, when the TP-AGB energetic contribution is maximum\cite{Maraston+05, Maraston+06}. This maximizes the contribution of the short TP-AGB phase ($\sim$3 Myr for an individual star) to the integrated galaxy SED, with respect to systems with continuing star formation and a lower mass fraction of stars in the appropriate age range. 

Indeed, we find that even for the best-performing M13 model, the $\chi_{\rm R}^2$ was large for the high SNR spectrum of D36123, which has stronger absorption features than the model. By separating optical data ($\lambda_{\rm rest}$$<$0.5$\mu$m) and NIR data ($\lambda_{\rm rest}$$>$0.5$\mu$m), we confirmed that the fit in the NIR spectrum leads to high values of $\chi^2$, as we found a minimum reduced $\rm \chi_R^2$$\approx$9-11 in the optical spectrum for the four different models (C09 performing best in this case), while the reduced $\chi^2_R$ remained large ($\rm \chi_R^2$$\approx$40-60) when fitting only the NIR spectrum (Table~\ref{tab1}). Interestingly, excluding from the fit the range $\lambda_{\rm rest}$$=$0.5-1.0$\mu$m, which contains the strongest TiO features, the M13 and M05 models largely perform better than the others when reproducing solely this region, pointing to a more consistent treatment of this spectral range, which is heavily affected by TP-AGB stars (Extended Data Fig.\ref{ext_fig02}). 

In any case, overall, $\rm \chi_R^2$ was always large for each model and spectral range, suggesting that improvements to the models are necessary. Focusing on the best-fitting models, M13 (like M05) uses empirical spectra for just the TP-AGB phase. For the other phases contributing to the NIR, as the RGB, classical Kurucz-type theoretical spectra from model atmospheres are adopted, which - as is well known - do not adequately match all NIR absorption features\cite{Baldwin+18, Dahmer-Hahn+18}. On the other hand, models based on the empirical RGB spectra perform better, but not for all features\cite{Baldwin+18}, as we also show here. Notably, various studies show that all CO NIR spectral features from H to K in local massive early-type galaxies (ETGs), whose NIR spectra are dominated by RGB stars, are systematically stronger than those in models based on empirical libraries\cite{Eftekhari+22}. This is also observed in the NIR spectra of local spirals and star-forming galaxies\cite{Riffel+19}. This evidence suggests that D36123 also contains stars that are more metal-rich than those included in the model. Moreover, the M13 models apply local samples of C-rich and O-rich spectra, which may not cover the range of metallicity and abundance ratios hosted in distant massive ellipticals. The metallicity dependence of the relative contributions of C-rich and O-rich TP-AGB stars in M13 models is included using a theoretical prescription\cite{Renzini+Voli+1981},  such as in more metal-rich models, more fuel consumption in O-rich rather than in C-rich phases, but they still use the same observed spectra. Moreover, the calibration of stellar population models requires samples of stellar generations that are as simple as possible, that is, star clusters, with independent age and chemical composition available from fitting a color-magnitude diagram and from stellar spectroscopy. For the TP-AGB phase, these calibrating clusters also need to span a wide age range, extending down to $\sim$100 Myr. Star clusters with these properties exist in the Magellanic Clouds and have traditionally served to calibrate stellar population models\cite{Maraston+1998, Conroy+10, Noel+13}. The limitation is that these clusters have subsolar metallicity and possibly solar or subsolar element ratios. 

Another issue occurs when there is a galaxy-specific chemical pattern. It is well-known that massive galaxies, both locally\cite{Thomas+05} as well as at high redshift\cite{Onodera+15, Lonoce+15}, feature [$\alpha$/Fe]-enhanced stellar populations. The same galaxies also have enhanced ratios of [C/Fe] and [N/Fe]\cite{Johansson+12, Worthey+14} and possibly also have Na\cite{Conroy+14, Parikh+18}. Empirical libraries based on local, solar-scaled RGB spectra, and the population models based on them, may not reflect these chemical patterns, causing discrepancies. This would mirror a long-standing discrepancy between optical spectral features of massive galaxies and early population models\cite{Worthey+1992}, which was solved by [$\alpha$/Fe]-enhanced population models\cite{Thomas+03}. Galaxy spectra like those we present here, with young ages, and supersolar metallicity and possibly various non-solar element enhancements, are unique, with unknown counterpart in the Universe. They challenge the models but also offer the invaluable opportunity to calibrate them, with far-reaching benefit also for stellar evolution and model atmosphere theories. 

Inferring the timeline of galaxy evolution depends on population synthesis models. In turn, the model age scale depends on the input stellar evolution in terms of stellar tracks (and their input) and the energetics of stellar phases. We derive systematically younger ages and lower masses when fitting these spectra with TP-AGB-mild models (M13) versus TP-AGB-poor/-light models (BC03/C09), especially at high redshift. The age differences range from $\sim$100 Myr to $\sim$1 Gyr (Tables~\ref{tab1} and \ref{tab2}). Younger galaxy ages can have implications for the inferred physics of galaxy evolution, such as gas accretion and star formation quenching. Also, galaxy simulations adopt evolutionary population synthesis models as input, hence our results impact the predicted spectra and photometry of simulated galaxies found in cosmological and other simulations\cite{Tonini+09}. In that case, the age was fixed within the simulation, but the simulated galaxy SEDs and the calibrations through comparisons to observations were obviously affected. The AGB stellar phase has also a major impact on galaxies' dust production and chemical evolution particularly for heavy elements\cite{Schneider+Maiolino+24}. More fuel for this phase, as supported by our findings, has implications on the relevant calculations for these processes\cite{Kelson+10}, thus potentially having an even broader impact on galaxy evolution.

The spectra presented here demonstrate the presence of TP-AGB spectral features, including those from carbon stars, in the integrated NIR light of high-redshift quiescent galaxies. We have observed these features in all their complexity in galaxy spectra. Clearly, our very small galaxy sample does not allow us to infer, statistically, the impact of the TP-AGB emission on the derived stellar population properties of young quiescent galaxies. Previous photometric studies\cite{Maraston+06, Capozzi+16} have shown that, on average, fitting distant galaxies using models with heavier TP-AGB contributions yields younger ages and lower stellar masses than TP-AGB-poor models, but with significant galaxy-by-galaxy differences. The usual degeneracies between age, reddening and metallicity still apply to some extent and, for individual galaxies, the fitting results for the SEDs can depend on unknown details of the past SFHs (for example). From our small sample, evaluating how frequently these TP-AGB features are present in early galaxy spectra is also difficult. NIRSpec observations of a larger, homogeneously selected sample of galaxies as luminous as D36123 (which are common, even over a single NIRSpec pointing) can address these questions and help us quantify the broader consequences of TP-AGB emission in determining galaxies' physical parameters. Such observations may also help guide further improvement of stellar population models.

\begin{small}
\begin{table*}[ht!]
\setlength{\tabcolsep}{13pt}
\centering
\caption{Stellar population properties of D36123.}\label{tab1}
\begin{tabular}{clllll} 
\hline
{\small Rest-frame}&Property & {\;  \;  } BC03& {\;  \; \;  }C09& {\;  \; \;  \;} M13& {\;  \; \;  \; }M05  \\
{\small spectral range} &  & TP-AGB-poor    & TP-AGB-light       & TP-AGB-mild          &TP-AGB-heavy          \\
\hline\hline
&   & \multicolumn{2}{c}{RA = 14h : 19m : 34.258s}&  \multicolumn{2}{c}{DEC = +52$^\circ$ : 56$^{\prime}$ : 23.079$\arcsec$} \\
\cline{3-6}
\vspace{3pt} 
&redshift &  1.074$^{+0.001}_{-0.001}$ & 1.075$^{+0.003}_{-0.001}$ & 1.082$^{+0.002}_{-0.002}$ &  1.075$^{+0.002}_{-0.004}$ \\
\vspace{3pt} 
&{$\rm M_*[\times 10^{10} M_{\odot}$]} & 1.493$^{+0.007}_{-0.004}$ & 1.489$^{+0.007}_{-0.006}$ &  1.167$^{+0.008}_{-0.005}$ & 1.552$^{+0.015}_{-0.014}$ \\
\vspace{3pt} 
&{\rm age [Gyr]} & 1.661$^{+0.024}_{-0.041}$  &  0.751$^{+0.025}_{-0.034}$ &  0.621$^{+0.029}_{-0.021}$ & 2.224$^{+0.016}_{-0.024}$ \\
\vspace{3pt} 
full&{\rm $Z/Z_\odot$} &  1.944$^{+0.028}_{-0.034}$  & 2.000$^{+0.000}_{-0.008}$ &  1.589$^{+0.061}_{-0.039}$ & 	0.982$^{+0.001}_{-0.001}$\\
\vspace{3pt} 
(0.3--2$\mu$m)&{\rm Av}  &  0.000$^{+0.002}_{-0.000}$ &  0.156$^{+0.016}_{-0.022}$  &  0.424$^{+0.026}_{-0.026}$  &	0.000$^{+0.000}_{-0.000}$ \\
\vspace{3pt} 
&{\rm $\tau$ [Gyr]}& 0.241$^{+0.019}_{-0.018}$ & 0.104$^{+0.014}_{-0.021}$  &  0.100$^{+0.010}_{-0.010}$  &  	0.707$^{+0.013}_{-0.017}$  \\  
\vspace{3pt} 
&SFR$_{\rm best}[\rm M_\odot/yr]$ & 0.294$^{+0.036}_{-0.065}$  &   0.099$^{+0.001}_{-0.001}$  & 0.311$^{+0.003}_{-0.003}$  &  1.320$^{+0.006}_{-0.006}$  \\
&SFR$_{\rm best}$/SFR$_{\rm peak}$ & 0.019$^{+0.011}_{-0.010}$  &   0.014$^{+0.022}_{-0.015}$  & 0.024$^{+0.024}_{-0.023}$  &  0.368$^{+0.032}_{-0.024}$  \\
&{\rm $\chi_{\rm R}^2$} & 59.6 & 52.9  & 39.0 & 102.6	\\   
\hline      
\vspace{3pt} 
&redshift & 1.078$^{+0.002}_{-0.002}$& 1.083$^{+0.001}_{-0.002}$ & 1.078$^{+0.001}_{-0.002}$ &  1.078$^{+0.001}_{-0.002}$ \\
\vspace{3pt} 
&{$\rm M_*[\times 10^{10} \rm M_{\odot}$]} &1.330$^{+0.019}_{-0.013}$ & 1.406$^{+0.033}_{-0.043}$ &  1.140$^{+0.011}_{-0.197}$ & 1.151$^{+0.024}_{-0.021}$ \\ 
\vspace{3pt} 
&{\rm age [Gyr]} & 0.961$^{+0.021}_{-0.026}$  &  0.810$^{+0.005}_{-0.003}$ &  0.909$^{+0.098}_{-0.098}$ & 0.969$^{+0.031}_{-0.069}$ \\
\vspace{3pt} 
optical &{\rm $Z/Z_\odot$} &  1.525$^{+0.095}_{-0.225}$  & 2.000$^{+0.000}_{-0.007}$ &  2.183$^{+0.017}_{-0.041}$ & 1.002$^{+0.048}_{-0.052}$\\
\vspace{3pt} 
($<0.5\mu$m)&{\rm Av}  & 0.068$^{+0.082}_{-0.068}$ &  0.000$^{+0.034}_{-0.000}$  &  0.000$^{+0.038}_{-0.000}$  &	0.000$^{+0.002}_{-0.000}$ \\
\vspace{3pt} 
&{\rm $\tau$ [Gyr]}& 0.007$^{+0.017}_{-0.006}$ & 0.091$^{+0.025}_{-0.026}$  & 0.001$^{+0.010}_{-0.000}$  & 0.007$^{+0.043}_{-0.000}$  \\ 
\vspace{3pt}
&SFR$_{\rm best}[\rm M_\odot/yr]$ & 0.000$^{+0.004}_{-0.000}$  &   0.093$^{+0.001}_{-0.001}$  & 0.000$^{+0.006}_{-0.000}$  &  0.000$^{+0.010}_{-0.000}$  \\
\vspace{3pt} 
&SFR$_{\rm best}$/SFR$_{\rm peak}$ & $\sim$0  &   $\sim$0  & $\sim$0  &  $\sim$0  \\
&{\rm $\chi_{\rm R}^2$} & 11.4 & 9.0  & 11.2 & 11.2\\ 
\hline      
\vspace{3pt} 
&redshift & 1.074$^{+0.001}_{-0.001}$& 1.080$^{+0.002}_{-0.004}$ & 1.083$^{+0.002}_{-0.002}$ &  1.080$^{+0.003}_{-0.001}$ \\
\vspace{3pt} 
&{$\rm M_*[\times 10^{10} \rm M_{\odot}$]} &1.679$^{+0.023}_{-0.031}$ & 1.514$^{+0.028}_{-0.021}$ &  1.247$^{+0.038}_{-0.017}$ & 0.556$^{+0.010}_{-0.004}$ \\ 
\vspace{3pt} 
&{\rm age [Gyr]} & 1.961$^{+0.011}_{-0.039}$  &  0.783$^{+0.031}_{-0.028}$ &  0.490$^{+0.041}_{-0.041}$ & 0.207$^{+0.014}_{-0.001}$ \\
\vspace{3pt} 
NIR &{\rm $Z/Z_\odot$} &  1.883$^{+0.040}_{-0.034}$  & 2.000$^{+0.000}_{-0.016}$ &  0.692$^{+0.029}_{-0.032}$ & 0.500$^{+0.022}_{-0.000}$\\
\vspace{3pt} 
($>0.5\mu$m)&{\rm Av}  & 0.000$^{+0.006}_{-0.000}$ &  0.103$^{+0.022}_{-0.022}$  &  0.706$^{+0.049}_{-0.044}$  &	0.982$^{+0.029}_{-0.012}$ \\
\vspace{3pt} 
&{\rm $\tau$ [Gyr]}& 0.497$^{+0.029}_{-0.032}$ & 0.057$^{+0.016}_{-0.017}$  & 0.023$^{+0.010}_{-0.010}$  & 2.605$^{+0.332}_{-0.316}$  \\  
\vspace{3pt} 
&SFR$_{\rm best} [\rm M_\odot/yr]$ & 0.984$^{+0.177}_{-0.148}$ &   0.000$^{+0.012}_{-0.000}$  & 0.000$^{+0.018}_{-0.000}$  &  23.725$^{+0.013}_{-0.104}$ \\
&SFR$_{\rm best}$/SFR$_{\rm peak}$ & 0.207$^{+0.057}_{-0.049}$  &   $\sim$0  & $\sim$0  &  0.200$^{+0.029}_{-0.024}$  \\\  
&{\rm $\chi_{\rm R}^2$} & 60.7 & 58.5 & 40.7 & 66.6	\\
\hline      
\vspace{3pt} 
&redshift & 1.076$^{+0.001}_{-0.001}$& 1.093$^{+0.001}_{-0.002}$ & 1.081$^{+0.001}_{-0.001}$ &  1.075$^{+0.002}_{-0.003}$ \\
\vspace{3pt} 
&{$\rm M_*[\times 10^{10} \rm M_{\odot}$]} &1.581$^{+0.015}_{-0.014}$ & 1.750$^{+0.087}_{-0.002}$ &  1.076$^{+0.008}_{-0.002}$ & 1.535$^{+0.021}_{-0.025}$ \\ 
\vspace{3pt} 
&{\rm age [Gyr]} & 1.627$^{+0.021}_{-0.019}$  &  1.446$^{+0.095}_{-0.046}$ &  0.712$^{+0.018}_{-0.012}$ & 2.183$^{+0.038}_{-0.040}$ \\
\vspace{3pt} 
full &{\rm $Z/Z_\odot$} &  2.300$^{+0.048}_{-0.036}$  & 0.300$^{+0.010}_{-0.009}$ &  1.285$^{+0.030}_{-0.023}$ & 0.990$^{+0.007}_{-0.007}$\\
\vspace{3pt} 
(excluding&{\rm Av}  & 0.000$^{+0.004}_{-0.000}$ &  0.078$^{+0.023}_{-0.017}$  &  0.068$^{+0.014}_{-0.013}$  &	0.000$^{+0.009}_{-0.000}$ \\
\vspace{3pt} 
0.5--1$\mu$m)&{\rm $\tau$ [Gyr]}& 0.028$^{+0.012}_{-0.008}$ & 0.127$^{+0.054}_{-0.065}$  & 0.095$^{+0.009}_{-0.010}$  & 0.690$^{+0.012}_{-0.014}$  \\  
\vspace{3pt} 
&SFR$_{\rm best} [\rm M_\odot/yr]$ & 0.000$^{+0.008}_{-0.000}$  &  0.000$^{+0.045}_{-0.000}$ & 0.084$^{+0.013}_{-0.012}$   &  1.306$^{+0.022}_{-0.010}$   \\
\vspace{3pt} 
&SFR$_{\rm best}$/SFR$_{peak}$ & $\sim$0  &   $\sim$0  & 0.011$^{+0.009}_{-0.008}$   &  0.363$^{+0.027}_{-0.022}$   \\
&{\rm $\chi_{\rm R}^2$} & 32.2 & 23.1  & 16.3 & 83.7	\\  
&{\rm $\chi_{\rm R, \rm msk}^2$} & 446.7 & 308.4  & 157.9 & 172.1\\ 
\hline
\end{tabular}
\begin{tablenotes}
\footnotesize 
\vspace{5pt}
\item The first column denotes the rest-frame range of wavelength adopted in the spectral fit. The third to sixth columns list the best-fitting physical properties derived by four models with different TP-AGB treatments. Specifically, the total stellar masses are based on the Chabrier\cite{Chabrier+03} IMF. Age means the mass-weighted age, that is., an average look-back time weighted by the SFH. The best-fitting parameters and errors are derived from the distribution of $\chi^2$, with the former obtained for the minimum $\chi^2$. The errors were determined after rescaling to $\chi^2_{\rm R} =1$.  SFR$_{\rm best}$/SFR$_{\rm peak}$ is a measure of quiescence, where the SFR$_{\rm best}$ is derived from the best fit by adopting delay-$\tau$ models, and SFR$_{\rm peak}$ corresponds to the SFR maximum at $t=\tau$. $\chi_{\rm R, msk}^2$ presents the reduced $\chi^2$ within the masked  0.5-1$\mu$m region during the spectral fit.
\end{tablenotes}
\end{table*}
\end{small}

\begin{small}
\begin{table*}
\setlength{\tabcolsep}{13pt}
\centering
\caption{Stellar population properties of galaxies 8595 and 9025.}\label{tab2}
\begin{tabular}{clllll} 
\hline
Target&Property & {\;  \; }BC03 & {\;  \;  \; }C09 & {\;  \;  \; \;}M13 & {\;  \;  \; \;}M05  \\
{\small (Rest-frame)} &  & TP-AGB-poor & TP-AGB-light& TP-AGB-mild &TP-AGB-heavy  \\
\hline\hline
& & \multicolumn{2}{c}{RA = 14h : 20m : 42.240s}&  \multicolumn{2}{c}{DEC = +53$^\circ$ : 01$^{\prime}$ : 50.298$\arcsec$} \\
\cline{3-6}
\vspace{3pt} 
& redshift  & 1.739$^{+0.004}_{-0.002}$  & 1.746$^{+0.002}_{-0.003}$   &  1.746$^{+0.002}_{-0.003}$  &  1.741$^{+0.002}_{-0.002}$ \\
\vspace{3pt} 
& {$\rm M_*[\times 10^{10} \rm M_{\odot}$]} & 0.791$^{+0.020}_{-0.020}$  &  0.736$^{+0.009}_{-0.007}$   &  0.545$^{+0.003}_{-0.003}$   &  0.676$^{+0.009}_{-0.009}$  \\ 
\vspace{3pt} 
&{\rm age [Gyr]} &1.913$^{+0.052}_{-0.052}$   &  0.798$^{+0.023}_{-0.023}$   &   0.732$^{+0.028}_{-0.027}$  &  1.908$^{+0.057}_{-0.058}$\\
\vspace{3pt} 
8595&{\rm $Z/Z_\odot$}  &1.711$^{+0.039}_{-0.039}$ &   2.000$^{+0.000}_{-0.006}$   &   1.584$^{+0.044}_{-0.046}$   &   1.878$^{+0.012}_{-0.012}$	\\
\vspace{3pt} 
{\small (full)}&{\rm Av}  & 0.012$^{+0.051}_{-0.012}$  &  0.357$^{+0.023}_{-0.030}$ & 0.368$^{+0.020}_{-0.021}$   &   0.013$^{+0.022}_{-0.013}$  \\
\vspace{3pt} 
&{\rm $\tau$ [Gyr]}& 0.283$^{+0.029}_{-0.034}$ & 0.134$^{+0.041}_{-0.048}$ &  0.133$^{+0.038}_{-0.055}$ &  0.403$^{+0.022}_{-0.025}$\\ 
\vspace{3pt} 
&SFR$_{\rm best} [\rm M_\odot/yr]$  & 0.073$^{+0.015}_{-0.015}$ &  0.049$^{+0.004}_{-0.004}$  & 0.061$^{+0.004}_{-0.004}$   &  0.232$^{+0.019}_{-0.018}$  \\
\vspace{3pt} 
&SFR$_{\rm best}$/SFR$_{\rm peak}$  & 0.021$^{+0.018}_{-0.015}$ &  0.042$^{+0.078}_{-0.078}$  & 0.061$^{+0.142}_{-0.098}$   &  0.113$^{+0.038}_{-0.034}$  \\
\vspace{3pt} 
&{\rm $\chi_{\rm R}^2$}& 0.821 & 0.950 & 0.794 & 0.935 \\   
\vspace{3pt} 
&{\rm $P(\Delta \chi^2$)}  & 0.00103   & 0 & -- & 0 \\  
\hline 
& &\multicolumn{2}{c}{RA = 14h : 19m : 33.991s}& \multicolumn{2}{c}{DEC = +52$^\circ$ : 51$^{\prime}$ : 56.290$\arcsec$} \\
\cline{3-6}
\vspace{3pt} 
&redshift &  1.545$^{+0.001}_{-0.001}$ & 1.546$^{+0.001}_{-0.002}$ & 1.569$^{+0.001}_{-0.002}$ &  1.569$^{+0.001}_{-0.002}$ \\
\vspace{3pt} 
&{$\rm M_*[\times 10^{10} \rm M_{\odot}$]} & 0.759$^{+0.082}_{-0.046}$   &  0.859$^{+0.087}_{-0.042}$  & 0.755$^{+0.052}_{-0.050}$    &  0.746$^{+0.025}_{-0.023}$	 \\ 
\vspace{3pt} 
&{\rm age [Gyr]} & 1.560$^{+0.073}_{-0.122}$ & 1.580$^{+0.065}_{-0.062}$ &  0.272$^{+0.045}_{-0.045}$ & 0.182$^{+0.001}_{-0.018}$ \\
\vspace{3pt} 
9025&{\rm $Z/Z_\odot$}& 2.123$^{+0.203}_{-0.075}$  &   0.300$^{+0.004}_{-0.003}$ &  1.700$^{+0.115}_{-0.124}$ &  1.140$^{+0.064}_{-0.045}$	 \\
\vspace{3pt} 
{\small (full+HST)}&{\rm Av} & 0.351$^{+0.061}_{-0.103}$ & 0.397$^{+0.026}_{-0.025}$ &  1.125$^{+0.048}_{-0.042}$ &1.337$^{+0.052}_{-0.046}$\\
\vspace{3pt} 
& {\rm $\tau$ [Gyr]} & 0.304$^{+0.037}_{-0.042}$  & 0.316$^{+0.063}_{-0.071}$    &  0.019$^{+0.008}_{-0.010}$ & 0.008$^{+0.006}_{-0.007}$	\\ 
\vspace{3pt} 
& SFR$_{\rm best} [\rm M_\odot/yr]$  & 0.181$^{+0.024}_{-0.013}$ &  0.230$^{+0.045}_{-0.028}$ & 0.048$^{+0.014}_{-0.014}$ &  0.027$^{+0.004}_{-0.004}$ \\
\vspace{3pt} 
& SFR$_{\rm best}$/SFR$_{\rm peak}$  & 0.082$^{+0.068}_{-0.056}$ &  0.092$^{+0.107}_{-0.095}$ & $\sim$0 &  $\sim$0  \\
\vspace{3pt} 
&{\rm $\chi_{\rm R}^2$}& 1.702 &1.838 & 1.634  & 1.649 \\  
\vspace{3pt} 
&{\rm $P(\Delta \chi^2 $)} & 0.00001 &  0  &  --  &  0.03767 \\
\hline
\end{tabular}
\begin{tablenotes}
\footnotesize 
\vspace{5pt}
\item For both objects, the spectral fit used the entire rest-frame wavelength range. Additionally, to better constrain the age-sensitive wavelength around 4000\AA, two other HST bands (F606W+F814W) were employed simultaneously in the fit of object 9025 (Methods). Measurements and errors are as in Table~\ref{tab1}, except that we did not rescale the reduced $\chi^2$ to 1. $\Delta \chi^2$ was computed relative to the M13 model. 
\end{tablenotes} 
\end{table*}
\end{small}

\clearpage
\newpage

\section*{Methods}
\subsection{NIRSpec observation and reduction\\}
In this study, we present NIR spectra for three quiescent galaxies, IDs=D36123, 8595, and 9025. We cross-matched the catalog in CANDELS EGS field\cite{Stefanon+17} with the NIRSpec detections in the DD-ERS-1345 (CEERS, PI: Steven Finkelstein) and DD-2075 (PI: Pablo Arrabal Haro) programs. Both D36123 and 8595 were observed with the PRISM\cite{Jakobsen+22}, spanning 0.60-5.30$\mu$m with varying spectral resolution R$\equiv \lambda/\Delta \lambda \approx$ 30 at $\lambda=$1.2$\mu$m to R$>300$ at $\lambda>$5$\mu$m. Object 9025 was observed with the G140M/F100LP, G235M/F170LP, and G395M/F290LP medium-resolution (R=500-1340) gratings\cite{Jakobsen+22}, covering 0.7-5.2$\mu$m. Three-shutter slitlets were employed, enabling a three-point nodding pattern to facilitate background subtraction. The total exposure time on sources 8595 and 9025 are the same at 3107s, while on source D36123 it is 18387s ($\sim$6$\times$ higher).

The NIRSpec data processing followed the same methodology as used for other CEERS NIRSpec observations\cite{Arrabal+Haro+23, Arrabal+Haro+23_natr, Fujimoto+23, Kocevski+23, Larson+23}. The NIRSpec data reduction was based on the STScI Calibration Pipeline\cite{Bushouse+2022} version 1.8.5 and the Calibration Reference Data System (CRDS) mapping 1041 for objects 8595 and 9025, and CRDS mapping 1029 for D36123. There are three main stages in the reduction process. Briefly, the {\tt calwebb\_detector1} pipeline module was utilized to correct for the detector 1/f noise, subtract the bias and dark current, and generate the count-rate maps (CRMs) from the uncalibrated images. There is an improvement in the {\tt jump} step to correct the cosmic ray ``snowball''\cite{Arrabal+Haro+23_natr}.  The stage two of the pipeline reduction is to create two-dimensional (2D) spectra with a rectified trace and flat slope from the generated CRMs. The background subtraction and slit loss correction of objects D36123 and 8595 were considered at this stage, while the slit loss correction ({\tt pathloss}) was not taken into account for object 9025 as it is poorly centered in the shutter. We verified that better overall $\chi^2_R$ of its spectral fit can be obtained if adopting the empirical aperture corrections directly, which is the last step of the absolute flux calibrations (as discussed later in a dedicated subsection) to complement any incorrect calibration during the reduction process. At the third stage, we made use of optimised apertures for the extraction of the one-dimensional (1D) spectra in all cases. For object D36123, we adopted a 4-pixel extraction aperture, considering the effects of the Micro-Shutter Assembly\cite{Ferruit+22} (MSA) bar shadows on some upper pixels in 2D spectra. We also considered 3- and 5-pixel apertures, with negligible impact on the results. The flux error for the extracted spectra was calculated by the JWST pipeline using an instrumental noise. We rescaled all flux errors multiplying by a factor of $\sim$1.5 to account for correlations induced by the pipeline\cite{Arrabal+Haro+23}.

The spectrum of 9025, observed with 3 medium-resolution gratings, was calibrated through several steps. We first matched the 3 spectra scaling them by a constant term, determined using the overlapping regions, with variance weighting. This required a scaling of factors of 1.23 and 1.28 between G140M and G235M and between G235M and G395M, respectively. In the overlapping regions, the final coadded spectra were obtained by combining their weighted spectra and errors, ensuring that standard errors in the weighted mean over common wavelength ranges remained unchanged after resampling. For the purpose of increasing the continuum SNR to display the NIR features and speeding up the spectral fit computation, we finally resampled this medium-resolution spectrum to the one in the PRISM resolution. The average flux and the standard error of the mean in each PRISM wavelength bin were adopted.

\subsection{Photometric measurements\\}
Two out of three galaxies (D36123 and 9025) were observed in CEERS/NIRCam\cite{Finkelstein+22} imaging, while all (including 8595) have CANDELS/HST\cite{Grogin+11, Koekemoer+11} imaging. We used {\tt galfit}\cite{Peng+02, Peng+10} to derive their structure and obtain photometry in NIRCam bands (such as F115W, F150W, F200W, F277W, F356W, F410W, and F444W, when available), in HST bands (such as F606W, F814W, F125W, F140W, F160W, when needed) and in CFHT/WIRCam Ks band (only for object 8595). We adopted 3\arcsec$\times$3\arcsec-wide cutouts with 0.03\arcsec  pix$^{-1}$ sampling in the NIRCam bands and 0.06\arcsec pix$^{-1}$ sampling in the HST bands. We fitted the galaxies by using single S{\'e}rsic models with free parameters, including the positions, S{\'e}rsic index ($n$), effective radius (r$\rm _e$), total magnitudes, axis ratio (b/a), and position angle (PA). The point spread functions (PSFs) in different NIRCam bands were adopted from previous CEERS work\cite{Gomez-Guijarro+23}. They were constructed by stacking point sources using the software {\bf \tt PSFEx}\cite{Bertin+11}. We carefully modelled all objects in the field of view, with special attention in the case of 8595 in the 5\arcsec$\times$5\arcsec region to a bright nearby galaxy. We took about 5\% systematic uncertainty of photometry into account for objects D36123 and 9025, while 10\% systematic uncertainty of photometry was adopted for object 8595 when considering the effect of the bright nearby galaxy. The structural parameters (D36123 and 9025 in F150W and 8595 in F160W) are listed on Extended Data Fig.1, that also shows the {\tt galfit} models and residuals.

\subsection{Aperture correction \\}
Spectra reduced with the NIRSpec pipeline include an aperture correction term that is derived assuming a point-source scenario. To account for the actual shapes of our galaxies, we integrate the observed spectra through imaging filter bandpasses to obtain synthetic photometry (i.e., $F_{syn}$), and compare to the actual photometry (i.e., $F_{phot}$) measured from imaging. We are implicitly assuming here that no strong color gradient biases the in-slit photometry with respect to the integrated galaxy photometry. This point has been verified by building PSF-matched color images of our targets and verifying that no color gradients were apparent. We thus derive the aperture correction as $R_{corr}= F_{phot}/F_{syn}$. The top panel of Supplementary Fig.1 shows the aperture correction for object D36123 derived in this way. We adopt a linear function to fit the trend with wavelength, as shown by the red line. There is a weak wavelength dependence (slope $\alpha \sim$0.02). The green line presents the rescaled aperture correction derived from the forward-modeling tool for NIRSpec Multi-Object Spectroscopic data, namely MSAFIT\cite{de+Graaff+23}, which accounts for the complex geometry of the target galaxy, point spread function and pixelation of the NIRSpec instrument. This MSAFIT line aligns with the red linear relation. Regardless of the relation (linear vs. MSAFIT) or when adopting the extremes in the $\pm$1$\sigma$ range in the linear relation (the grey shaded region of D36123 in Supplementary Fig.1) to correct the spectrum, the best-fit parameters and spectral fits show minimal changes. Hence we decided to finally rescale the observed spectrum of D36123 by adopting the linear relation. The same process is applied to the other two objects 8595 and 9025. Their spectra are rescaled multiplying by a constant value of $\sim$2.28 and a linear relation with a slope of $\sim$ -0.07, respectively. 

\subsection{Spectral resolution\\}
When comparing the observed spectra with spectral synthesis models we need to match their spectral resolution. For the observed spectra, we have to consider the instrumental broadening, as well as the intrinsic linewidths from the galaxies. For the template models, we need to estimate their intrinsic spectral resolution. We will discuss in turn these terms in the following. 

The spectral resolution of the NIRSpec PRISM is relatively low. Based on the spectral sampling and for an unresolved source filling the slit, we can expect a resolution of 70 at 0.6$\mu$m, declining to 30 at 1.1$\mu$m, and rising again to 200 at 5$\mu$m. For point/compact sources, the resolution of in-flight data could be 1.5-2.0$\times$ higher than that in the pre-launch approximation\cite{de+Graaff+23}. This would be helpful to identify some weak/narrower features, especially for galaxy D36123 where the SNR in the continuum is high. We adopt the MSAFIT\cite{de+Graaff+23} software to model the NIRSpec resolving power based on the actual galaxy spatial profile. The resolution modelled by MSAFIT software is consistent with that of a flat source in a slit width of 1.4 pixels (see green curve in the top-middle panel of Supplementary Fig.2).

The intrinsic dispersion of our target early-type galaxies (ETGs) needs to be accounted next. The velocity dispersion $\sigma$ can be estimated using local stellar mass vs. velocity dispersion relations (i.e., $M_*$-$\sigma$) relation\cite{Thomas+05}. We convert $\sigma$ to the full width at half maximum (FWHM) and define $1/R=v_{\rm (rest, FWHM)}/c=\Delta\lambda/\lambda$ (where $\Delta\lambda=$FWHM). The intrinsic broadening of an ETG with a velocity dispersion $\sim$130km/s (a ballpark number for the typical masses $\sim10^{10}M_\odot$ in our study) corresponds to R$\sim$1000 (see the grey line in the top-middle panel of Supplementary Fig.2), quite negligible compared to the instrumental spread.

Finally, we estimated the template model's resolution from their spectral sampling, adopting 2 spectral bin widths as their (Nyquist) sampling. The sampling (hence resolution) is changing in steps, higher in the optical than in the NIR. Redshifting will affect where these steps actually apply for the observed galaxies. The corresponding resolution in the observed frame for D36123 (using {$z=$1.08}), is shown in the top-middle panel of Supplementary Fig.2. 

To match the resolution of the models with the observed spectrum, we need to account for all three impacting factors: prism dispersion (dominant term), the wavelength sampling of the spectral template models, and the velocity dispersion of the ETG. We add and subtract in quadrature the corresponding terms at fixed $\lambda_{\rm obs}$. The broadening $\Delta\lambda_{\rm fit}$ needed to be applied to models to match the observed spectra is provided by the equation: 
\begin{equation}
 \rm \Delta \lambda_{fit}=\sqrt{\Delta \lambda_{PRISM(1.4\;pix)}^2-\Delta \lambda_{MOD_i}^2+\Delta \lambda_{ETG}^2} ,
\label{eq1}
\end{equation}
where $i$ refers to the model. The final smoothing functions applied to models are equivalent to resolutions ranging from R$\sim$ 50 to 400 across the observed wavelength range of 0.6--3.3$\mu$m, as depicted in the top right panel of Supplementary Fig.2. Beyond $\sim$3$\mu$m for some models we are limited by the models' resolution. No further smoothing is applied to those. We repeat a similar procedures to obtain the appropriate smoothing kernels for objects 8595 and 9025, shown in the middle and bottom panels of Supplementary Fig.2. For both, the prism resolution is based on 1.4-pixel elements, and the spectra in different models were resampled in every 2 spectral bin widths, similar to what is done for object D36123.

\subsection{Spectral fit \\}
To estimate the stellar mass, age, and SFH of a galaxy, we adopt a custom IDL routine\cite{Gobat+12} to fit the final corrected 1D spectrum of each galaxy. The best-fitting results are derived by comparing the 1D spectrum with a range of composite stellar population templates by $\chi^2$-minimisation. These templates were generated by combining a grid of different SSP models assuming a delayed exponentially declining SFH ($\propto (t/\tau^2)e^{-t/\tau}$, i.e., delayed-$\tau$ model). The characteristic timescale $\tau$ and age variations (i.e., the time $t$ since the onset of star formation) are the same in different models, shown in Supplementary Table 2. To identify the redshift, the Calzetti attenuation law\cite{Calzetti+2000} was adopted, and the stellar metallicity was left free with lower and upper limits dependent on the model. To reduce the computational cost, we run the spectral fit twice. First, we run the spectral fit in a large range of redshift with a low-resolution grid ($\Delta z=0.1$). After the most probable redshift peak was identified, we narrowed down the redshift range with a high-resolution grid ($\Delta z= 0.001$) in the second run. The redshift boundary in Supplementary Table 2 is an example of a parametric grid for D36123 in the second run. For object D36123 with the highest SNR, the spectral fit was repeated varying the wavelength range (see Table 1): (1) with the full rest-frame wavelength (0.3-2$\mu$m, see Extended Data Fig.2a-d); (2) with the optical only ($\lambda_{\rm rest} <0.5 \mu$m); (3) with the NIR only ($\lambda_{\rm rest} >0.5 \mu$m); and (4) with the full rest-frame wavelength range but excluding 0.5--1$\mu$m (see Extended Data Fig.2e-h). For the two fainter galaxies, the spectral fit was only done in the full rest-frame wavelength. Since the medium resolution grating spectrum of object 9025 does not cover too well the age-sensitive spectral region around rest-frame 4000$\rm \AA$ break, we additionally considered the photometry in HST F606W and F814W bands to better constrain the overall fit. 

The best-fitting parameters with 1$\sigma$ errors\cite{Avni+1976} can be derived from the distribution of $\chi^2$, which are listed in Table~2. Considering the large minimum reduced $\chi^2$ of object D36123, we derived the uncertainties (as listed in Table~1) scaling up all errors to obtain a $\chi^2_R \sim$1.0. Stellar masses are reported for a Chabrier IMF~\cite{Chabrier+03} and are corrected for stellar losses and include remnants. Ages reported in Tables and discussed in the paper are mass-weighted.
Extended Data Fig.2 shows the best-fit model for each galaxy using models in red at the top of each panel, and the relative flux deviation $\rm D_{REL}=(F_{\lambda, obs}-F_{\lambda, mod}) / F_{\lambda, obs}$ at the bottom of each panel. The SNR-weighted averages of the relative deviations are labelled at the top of each bottom panel. For object D36123 the models deviate from the spectrum at a few percent levels (2.5\% level for M13, and higher values for the other models). This is likely driven by systematics in the models (plus, possible residual calibration errors) which we can detect at the high SNR. Higher deviations correspond to observed features not matched by the models. For the other galaxies, the relative flux deviations are larger and dominated by the noise in the data.

{\bf Fitting additional models to D36123:}
In addition to the models discussed in the main text, we also fitted three additional models to the galaxy D36123. This includes two fully empirical sets of models as in XSL\cite{Verro+22} and E-MILES\cite{Vazdekis+15}. We also tested CB07 models (https://www.bruzual.org/cb07/), an update of BC03 that was meant to include strong TP-AGB (never formally published). Some descriptions of the recipes of these models are provided in a further section in these Methods. Their intrinsic dispersion and resolutions are also shown on the top panels of Supplementary Fig.2. Both XSL and E-MILES models have higher intrinsic resolutions than other models. The parameter grid settings in these three models are the same as those of the four main models, except for the metallicity due to available ranges (see Supplementary Table 2 for details). Their corresponding best-fit models are displayed in Extended Data Fig.3, accompanied by the best-fit redshift, mass-weighted age, stellar mass and reduced $\rm \chi_R^2$ shown in Extended Data Table 2, only for the full-range fits. The XSL model does not cover the blueward of $\lambda_{\rm rest}=$3500\AA, and includes two noisy regions corresponding to the low atmospheric transmission which are masked during the spectral fit (see grey-shaded regions of Extended Data Fig.3).

\subsection{Selected features and indexes on D36123\\}
We provide quantitative measurements of a range of features and indexes in D36123, chosen to be particularly representative as a test of the underlying contributing stellar phases, and also of models. To this aim, we have to take into account the fairly reduced spectral resolution of the PRISM spectrum of D36123, which triggers ad-hoc re-definition of  some indexes even when similar indexes were already defined in the literature. However by measuring these indexes in the same way in matched-resolution best-fitting models as well in representative stellar phases (including C- and O-type TP-AGB\cite{Lancon+2000, Lancon+02} and RGB\cite{Verro+22}) also at matched resolutions, we can gain valuable scientific insights (see Fig.3).

We define two EW indexes, that are estimated with the equation:
\begin{equation}
\rm EW(\lambda) = \int_{\lambda_1}^{\lambda_2} \frac{F_{c}-F_\lambda}{F_{c}} d\lambda,
\label{eq2}
\end{equation}
where F$_{\rm c}$ is the continuum obtained by drawing a straight line (see the blue dashed lines in Fig.3a,b) between the two-side average continuum from $\lambda_1$ to $\lambda_2$, estimated over 100\AA\ (rest) regions (red crosses). \\
1) EW(0.5-0.53$\mu$m) index: this includes MgI and MgII features, overlapping the MgII index, but also TiO and CN features. The continuum side windows are at the rest-frame 0.5-0.51$\mu$m and 0.53-0.54$\mu$m, while the index is integrated over rest 0.51-0.53$\mu$m. \\
2) EW(0.9-1.0$\mu$m) index: this covers the TiO+ZrO `valley' and the CN0.92 edge. The continuum side windows are at the rest-frame 0.91-0.92$\mu$m and 0.98-0.99$\mu$m, while the index is integrated over rest 0.92-0.986$\mu$m.

We further define two strength indexes as follows:\\
3) H-bump index. This broad bump, sensitive to the presence of cool stars, is the combination of the H$^-$ peak opacity at 1.6$\mu$m and of H$_2$O vapour absorptions around 1.4$\mu$m and 1.9$\mu$m. We adopt the literature  definition\cite{Worthey+1994}:
\begin{equation}
\rm I(H{\text -}bump) = -2.5 \log [ (\frac{1}{\lambda_2 - \lambda_1}) \int_{\lambda_1}^{\lambda_2} \frac{F_\lambda}{F_{cont}} d\lambda ],
\label{eq3}
\end{equation}
the F$_{\rm cont}$ pseudo-continuum flux (see the blue dashed line in Fig.3c) is estimated by connecting the midpoint of the two-side continuum windows at rest-frame 1.45-1.47$\mu$m and 1.765-1.785$\mu$m (red crosses). F$_\lambda$ is integrated from $\lambda_1=$1.61 $\mu$m to $\lambda_2=$1.67$\mu$m.\\
4) CN1.1 index. Among all CN edges, CN1.1 is the strongest detection in D36123. We define the CN1.1 index by the following formula:
\begin{equation}
\rm I(CN1.1) = -2.5 \log < F^{'}_{cont} / F_\lambda > ,  
\label{eq4}
\end{equation}
where the pseudo-continuum $\rm F^{'}_{cont}$ has a normalization and slope that are both measured using a single window at rest wavelength bluer than the edge between $\sim$1.05-1.075$\mu$m, $\sim$250\AA\ wide. The need to define a slope is to allow distinguishing steeply declining continuum over a large range from genuine edge-like discontinuities. A positive (rising slope) is, however, reset to a flat continuum to avoid inflating artificially the index. We then extrapolate this pseudo-continuum over the range $\sim$1.1-1.12$\mu$m where we integrate the spectrum (see Fig.3d).

Index measurements for D36123, best-fitting models and stellar templates of interest are shown in Fig.3. All the measurements are also listed in Extended Data Table 1. Measurements from models have obviously no associated uncertainties.

\subsection{Further details on model input\\}
The BC03 models\cite{Bruzual+Charlot+03} are calculated with an \textit{isochrone synthesis} algorithm, using Padova and Geneva tracks, plus extra ingredients to describe the TP-AGB phase and adopt both theoretical (BaSeL) and observational (STELIB and Pickles) libraries of stellar spectra, where the STELIB and Pickles libraries are extended to the shorter and longer wavelengths by using the BaSeL. The CB07 (and CB16) models (https://www.bruzual.org/) are updated versions of the BC03 model. The C09 models allow several user options and have been modified through the years. The models used in this work were generated using the latest FSPS version 3.2 by adopting a Chabrier IMF, the BaSTI isochrones and the MILES stellar library in the optical. The C09 models include TP-AGB spectra, extrapolated blueward with simpler linear slopes (as advocated by Lancon \& Wood 2000\cite{Lancon+2000}) and redward with the Aringer et al. (2009)\cite{Aringer+09} synthetic carbon star spectral library. The spectra of the oxygen-rich TP-AGB stars are extrapolated with the latest version of the PHOENIX stellar spectral library (the BT-SETTL library). For the TP-AGB isochrones, there is no shift adopted in the $\log$($L_{\rm bol}$) vs. $\log$($T_{\rm eff}$). Maraston models (i.e., M05 and M13) utilise the \textit{fuel-consumption theorem} for the treatment of post-main-sequence phases\cite{Maraston+1998} and adopt empirical C-/O-rich spectra for the TP-AGB phase\cite{Lancon+02} and the BaSeL library for the other phases. The XSL model\cite{Verro+22} utilises the PARSEC/COLIBRI isochrones including a TP-AGB contribution and adopting the X-shooter Spectral Library, which contains many evolved cool giants, covering the wavelength range from 3500$\rm \AA$ to 2.48$\mu$m (shorter than other models). These cool giants with $T_{\rm eff}<$4000 K are incorporated into the XSL model by using average spectra of static giants (i.e., RGB), O-/C-rich TP-AGB stars, binned by broad-band colour in the Vega system. The E-MILES model\cite{Vazdekis+15} adopts BaSTI stellar tracks for all phases, including the TP-AGB and the E-MILES library (http://miles.iac.es/pages/) for stellar spectra. All models assume a Chabrier IMF and are corrected to a solar metallicity $Z_\odot=0.02$ in this work. For the crude understanding of the effect of TP-AGB stars in the NIR mass-to-light ratio, we show $\sim$1 Gyr SSP models around solar metallicity from the different libraries in Supplementary Fig.3. SSPs from CB16, Vazdekis, XSL, E-MILES, and CB07 are degraded to match the wavelength-dependent resolution of M13 for better visualization.

\captionsetup[table]{name=Extended Data Table}
\captionsetup[figure]{name=Extended Data Fig.}
\setcounter{figure}{0}    
\setcounter{table}{0}

\begin{figure*}[htbp]
\begin{center}
\includegraphics[width=0.90\linewidth]{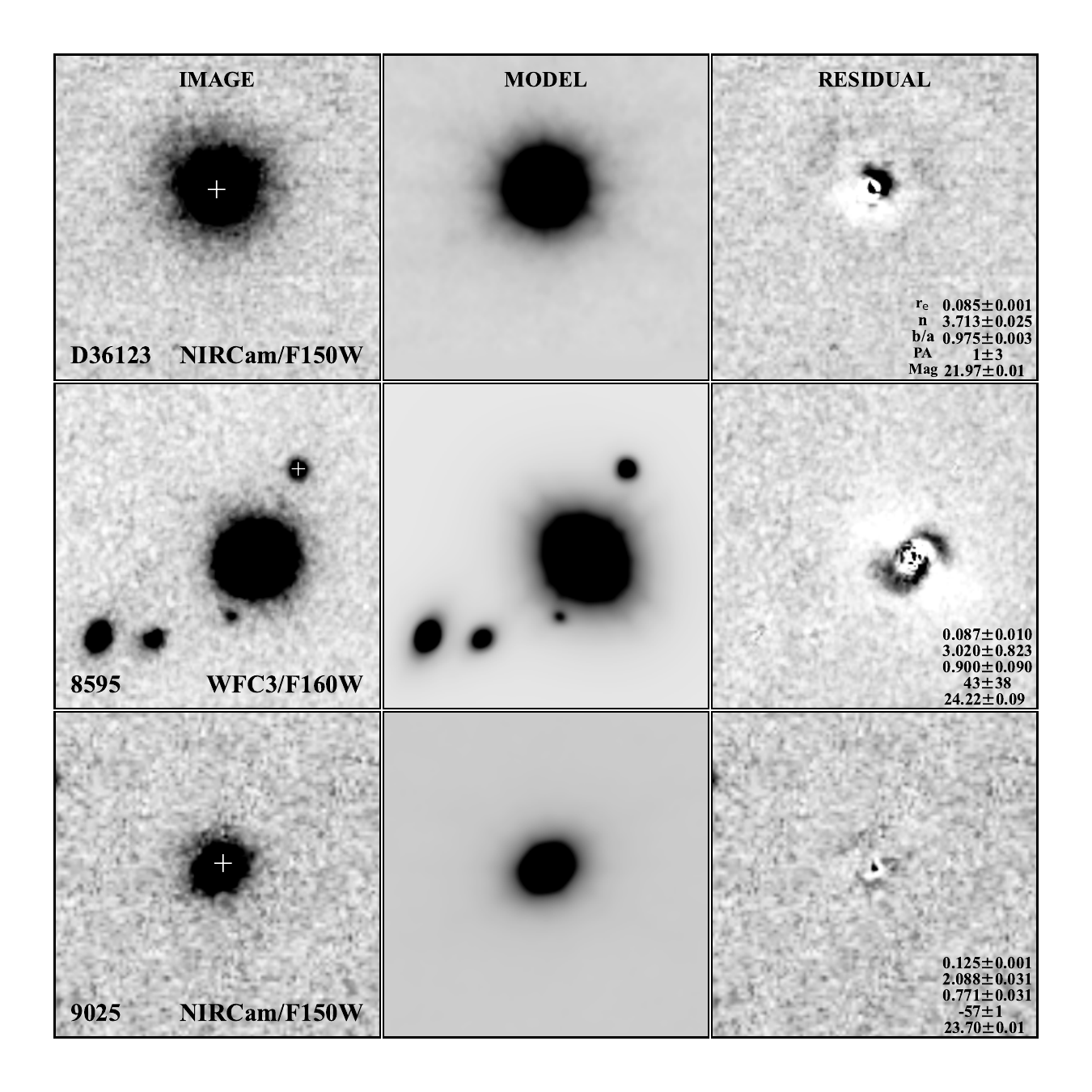}
\end{center}
\vspace{-0.3truecm}  
\caption{\small \textbf{Example of {\tt galfit} results.} The original image, best-fit model, and residual of object D36123 in NIRCam/F150W are shown in the top row. The results of objects 8595 in WFC3/F160W and 9025 in NIRCam/F150W are presented in the middle and bottom rows, respectively. The three targets are marked by white crosses on the original images (notice that 8595 is not at the center of the cutout). The cutouts of objects D36123 and 9025 are 3\arcsec$\times$3\arcsec wide, with a scale of 0.03\arcsec pix$^{-1}$. The cutouts for object 8595 are  5\arcsec$\times$5\arcsec wide with a pixel scale of 0.06\arcsec pix$^{-1}$. The best-fit structural parameters of each object are shown at the bottom of each residual panel, including the effective radius (r$\rm _e$) in arcsec, S{\'e}rsic index ($n$), axis ratio (b/a), position angle (PA) in degree, and AB Magnitude.} \label{ext_fig01}
\vspace{-12pt}
\end{figure*}

\begin{figure*}[htbp]
\begin{center}
\includegraphics[width=0.95\linewidth]{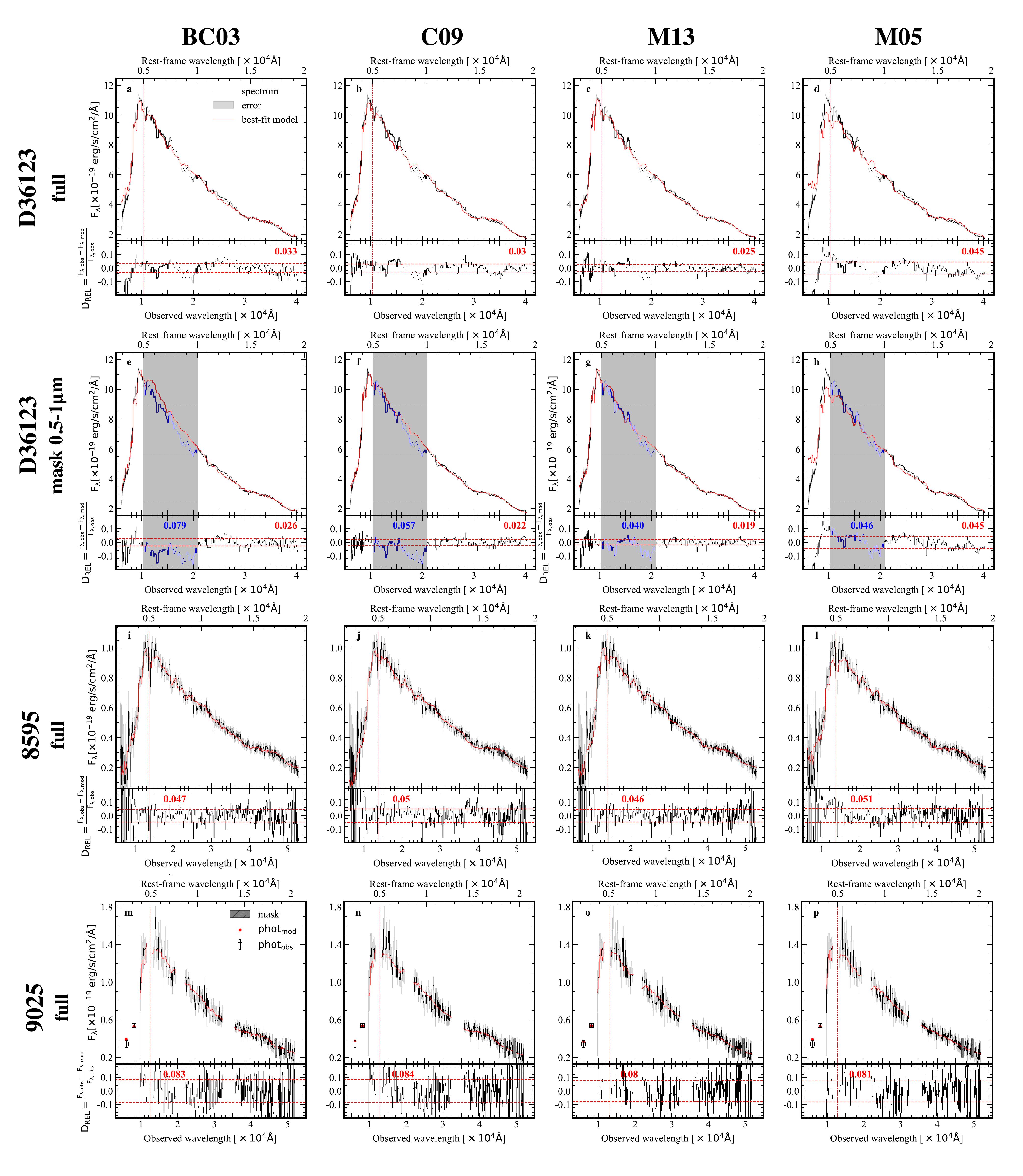}
\end{center}
\vspace{-0.3truecm}  
\caption{\small \textbf{Best-fit results and relative deviations}. From left to right, the best fits (top) and relative deviations (bottom) obtained by fitting BC03, C09, M13, and M05 models are shown. The panels from top to bottom present the results corresponding to D36123 in the full fit (1st) and in masking 0.5-1$\mu$m fit (2nd), 8595 (3rd) and 9025 (4th) in the full fit, respectively. The red horizontal dashed lines are the SNR-weighted mean of the relative deviation in the fitted region, and the corresponding value within the fitted (masked) region is printed in red (blue) on the top of each bottom panel. Detector defects in the spectrum of object 9025 are masked by grey rectangles.}\label{ext_fig02}
\vspace{-12pt}
\end{figure*}

\begin{figure*}[htbp]
\begin{center}
\includegraphics[width=0.95\linewidth]{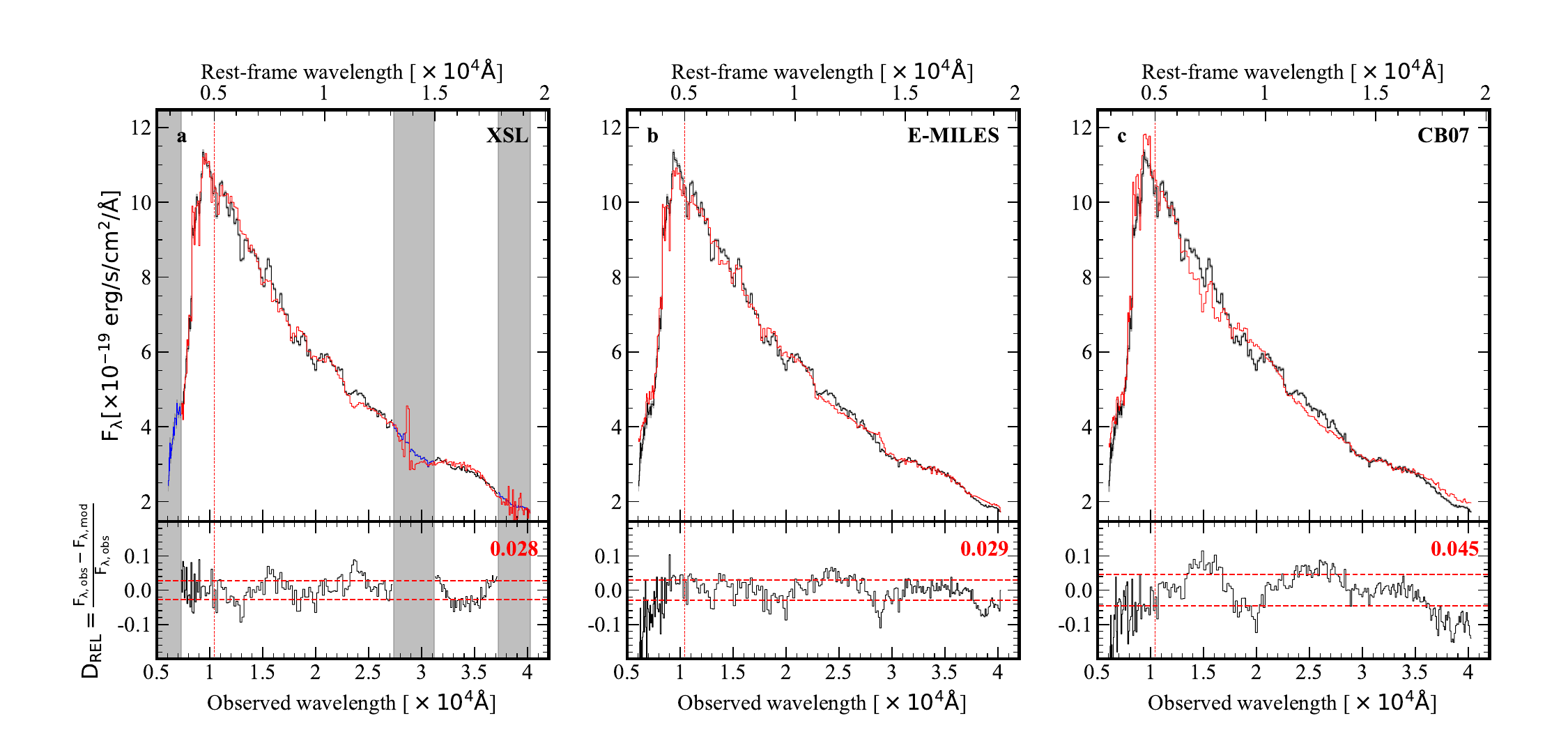}
\end{center}
\vspace{-0.3truecm}  
\caption{\small \textbf{Best-fitting spectra and relative deviations for D36123 based on  XSL, E-MILES, and CB07 models.}Analogous to the previous figure. The grey-shaded areas in the XSL model show the uncovered region at a short wavelength and two noisy telluric regions at a longer wavelength. }\label{ext_fig03}
\vspace{-12pt}
\end{figure*}

\begin{small}
\begin{table*}
\setlength{\tabcolsep}{2pt}
\centering
\caption{\textcolor{black}{Selected indexes and features quantitatively measured.}}\label{ext_tab1}
\begin{tabular}{c|c|c|c|c|c} 
\hline
&   & H-bump index & CN1.1 index & EW(Mg+TiO+C$_2$) & EW(CN0.92+TiO+ZrO)     \\
Spec&Bin& 1.61-1.67$\mu$m & $\sim$1.1$\mu$m & 0.5-0.53$\mu$m & 0.9-1$\mu$m  \\
\hline
D36123& -- &-0.084$\pm$0.002 & -0.116$\pm$0.003 & 9.639$\pm$0.486 & 33.569$\pm$0.944 \\
\hline
BC03 &  -- &-0.048 & -0.014$\pm$0.027 & 5.246 & 6.934   \\
C09 &  -- &-0.143  & -0.036$\pm$0.019 & 4.933 & 12.716  \\
M13 & --&-0.115  & -0.051$\pm$0.009 & 3.908 & 11.311  \\
M05 & --&-0.147  & -0.077$\pm$0.004 & 5.841 & 11.009 \\
\hline
XSL  & --  & -0.158 &0.007$\pm$0.001 & 6.793 & 19.967\\
E-MILES&-- & -0.081 &-0.011$\pm$0.004& 3.369 & 16.151\\
CB07  &--  & -0.028 &-0.034$\pm$0.008& 3.462 & 8.453\\
\hline
        &(I-K)=2.01&-0.124  &-0.014  &20.466  &12.771  \\
RGB$^{a}$ &(I-K)=2.54&-0.144  &-0.043  &22.407  &22.139  \\
        &(I-K)=3.41&-0.143  &-0.031  &23.838  &30.149  \\
\hline
                 &(I-K)=2.11 &-0.304 &-0.030  & 22.849 & 28.061 \\
O-rich TP-AGB$^{a}$&(I-K)=3.19 &-0.378 &-0.112  & 23.624 & 47.078 \\
                 &(I-K)=4.88 &-0.690 &-0.105  & -- & -- \\
\hline
                 &bin1 &-0.281 &-0.061  & -- & -- \\
O-rich TP-AGB$^{b}$&bin5 &-0.370 &-0.01  & -- & -- \\
\hline
                 &(I-K)=2.61 &-0.218 &-0.549  &43.837  &228.035  \\
C-rich TP-AGB$^{a}$&(I-K)=3.65 &-0.408 &-0.498  &72.194  & 257.859  \\
                 &(I-K)=4.46 &-0.491 &-0.358  &--  & -- \\
\hline
                 &bin1 &-0.338 &-0.385  &--  &--  \\
C-rich TP-AGB$^{b}$&bin3 &-0.299 &-0.309  &--  & --  \\
\hline
\end{tabular}
\begin{tablenotes}
\footnotesize 
\vspace{5pt}
\item \textcolor{black}{The $^a$ represents the average spectra of cool giants binned by broad-band colour (in Vega system) from the XSL simple stellar population models\cite{Verro+22}, while $^b$ are binned O- and C-rich spectra, observed by Lan{\c{c}}on \& Wood\cite{Lancon+2000}, and Lan{\c{c}}con \& Mouhcine\cite{Lancon+02}, respectively. Each feature of star spectra is measured after smoothing and rebinning to match the PRISM resolution around the feature position.} 
\end{tablenotes}
\end{table*}
\end{small}

\begin{small}
\begin{table*}[ht!]
\setlength{\tabcolsep}{13pt}
\centering
\caption{\textcolor{black}{Stellar population properties of D36123 modeled based on XSL, E-MILES and CB07.}}\label{ext_tab2}
\begin{tabular}{cllll} 
\hline
{\small Rest-frame}&Property & {\;  \;  } XSL& {\;  \; \;  }E-MILES& {\;  \; \;  \;} CB07  \\
\hline\hline
\vspace{3pt} 
&redshift & 1.084$^{+0.001}_{-0.001}$ &1.083$^{+0.001}_{-0.001}$ & 1.080$^{+0.001}_{-0.001}$ \\
\vspace{3pt} 
&{$\rm M_*[\times 10^{10} M_{\odot}$]} &1.866$^{+0.022}_{-0.022}$ &2.065$^{+0.029}_{-0.038}$ & 1.119$^{+0.009}_{-0.008}$  \\
\vspace{3pt} 
&{\rm age [Gyr]} &1.556$^{+0.047}_{-0.114}$  & 1.934$^{+0.026}_{-0.024}$  &  0.991$^{+0.001}_{-0.001}$\\
\vspace{3pt} 
full&{\rm $Z/Z_\odot$} & 1.319$^{+0.041}_{-0.034}$  &0.506$^{+0.011}_{-0.012}$  & 0.975$^{+0.004}_{-0.003}$  \\
\vspace{3pt} 
(0.3--2$\mu$m)&{\rm Av}  & 0.052$^{+0.023}_{-0.024}$ & 0.000$^{+0.003}_{-0.000}$   & 0.975$^{+0.004}_{-0.003}$   \\
\vspace{3pt} 
&{\rm $\tau$ [Gyr]}& 0.159$^{+0.034}_{-0.031}$ & 0.385$^{+0.023}_{-0.027}$  & 0.007$^{+0.028}_{-0.007}$     \\  
\vspace{3pt} 
&SFR$_{best}[\rm M_\odot/yr]$ & 0.073$^{+0.045}_{-0.047}$   & 0.460$^{+0.023}_{-0.023}$  & 0.000$^{+0.003}_{-0.000}$  \\
&SFR$_{best}$/SFR$_{peak}$ & 0.001$^{+0.003}_{-0.003}$ & 0.090$^{+0.033}_{-0.028}$  & $\sim$0  \\
&{\rm $\chi_{R}^2$} &55.5  & 46.1  & 117.0 \\  
\hline    
\end{tabular}
\begin{tablenotes}
\footnotesize 
\vspace{5pt}
\item[]{Measurements and errors are as in Table~\ref{tab1}. } 
\end{tablenotes}
\end{table*}
\end{small}

\captionsetup[table]{name=Supplementary Table}
\captionsetup[figure]{name=Supplementary Fig.}
\setcounter{figure}{0}    
\setcounter{table}{0}

\begin{figure*}[htbp]
\begin{center}
\includegraphics[width=0.6\linewidth]{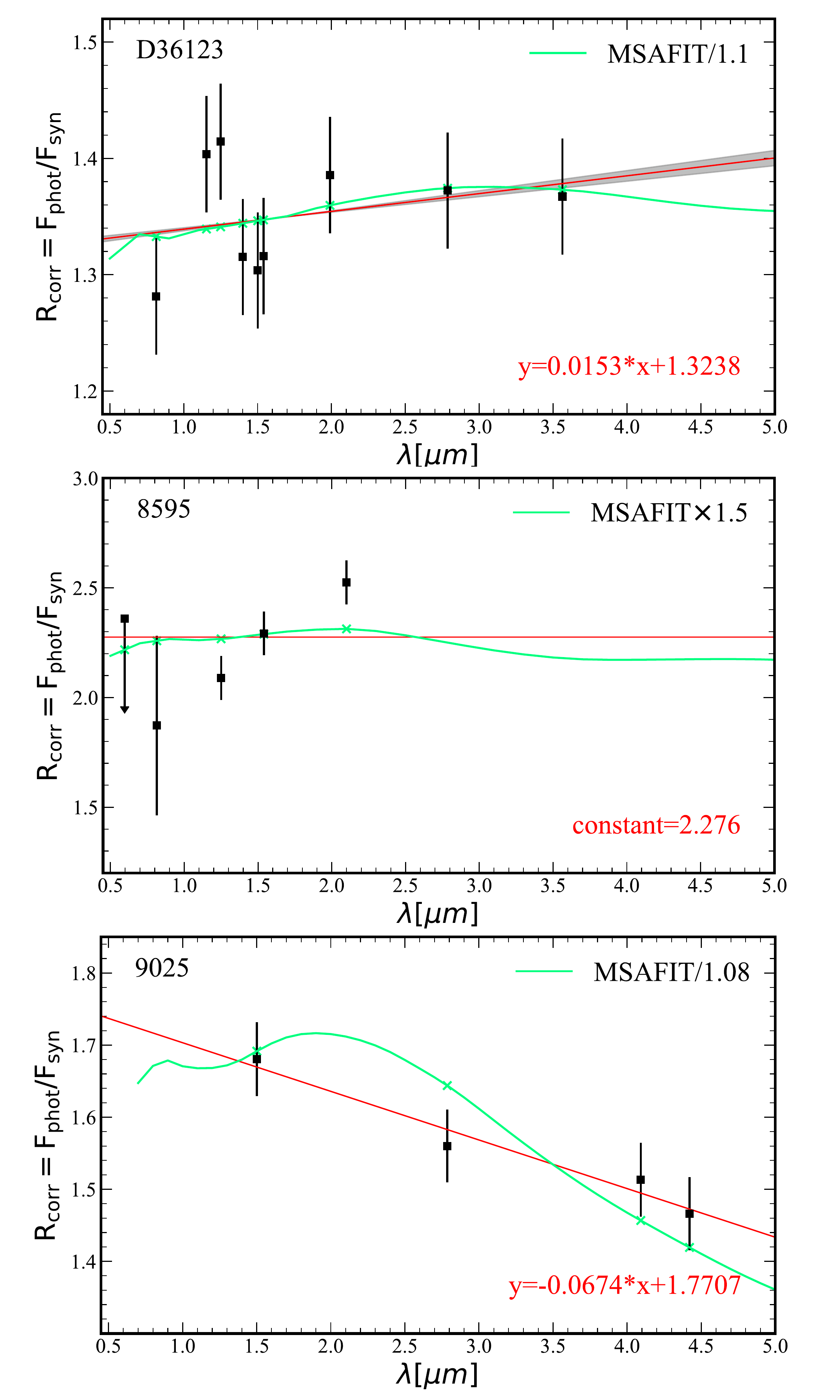}
\end{center}
\vspace{-0.3truecm}  
\caption{\small \textbf{Aperture correction.} Aperture correction ($R_{corr}$) in different bands as obtained through the ratio of the measured imaging photometry ($F_{phot}$) to the synthetic photometry ($F_{syn}$). The latter is obtained by integrating the spectra through the filter bandpasses. Panels from top to bottom show aperture corrections for objects D36123, 8595, and 9025. The aperture corrections adopted for the reported analysis are shown in red (after checking that no meaningful change follow from adopting the alternative shapes). The gray shaded region for object D36123 (which has a much higher SNR) presents the uncertainty on the linear aperture correction relation. The adoption of relations within the $\pm1\sigma$ range also lead to no significant change in the result. The green curves are derived by the MSAFIT\cite{de+Graaff+23} code, based on the measured size and S{\'e}rsic profiles for our targets. } \label{SI_fig01}
\vspace{-12pt}
\end{figure*}

\begin{figure*}[htbp]
\begin{center}
\includegraphics[width=0.95\linewidth]{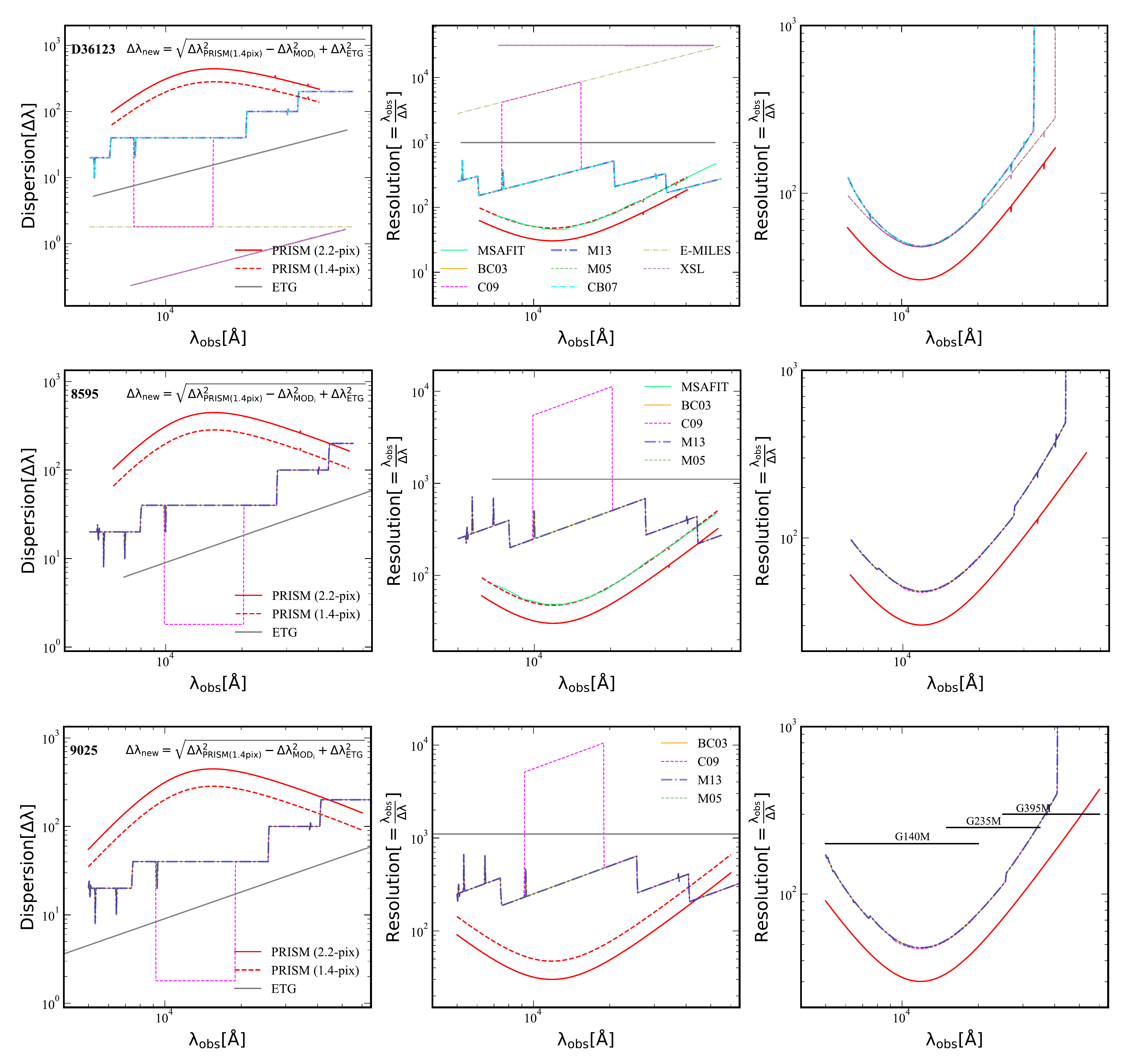}
\end{center}
\vspace{-0.3truecm}  
\caption{\small \textbf{Accounting for spectral resolution effects in matching observed spectra and spectral synthesis models.} The left panels display the dispersions $\Delta \lambda$ [$\rm \AA$], the middle panels present the corresponding resolutions, while the right panels show the effective resolution for convolving models to match that of the observed spectra. Instrumental resolutions for the PRISM based on 2.2-/1.4-pix elements in the shutter are in solid/dashed red lines. The instrumental resolution calculated by the MSAFIT code\cite{de+Graaff+23} based on actual galaxy sizes and shapes is shown in green for objects D36123 and 8595, except for object 9025 where the spectrum is resampled to the PRISM resolution. The model resolutions are in different colors. The intrinsic dispersions (converted to resolution) of the galaxies, based on local ETGs of the same stellar mass, are in grey.}
\label{SI_fig02}
\vspace{-12pt}
\end{figure*}

\begin{figure*}[htbp]
\begin{center}
\includegraphics[width=0.95\linewidth]{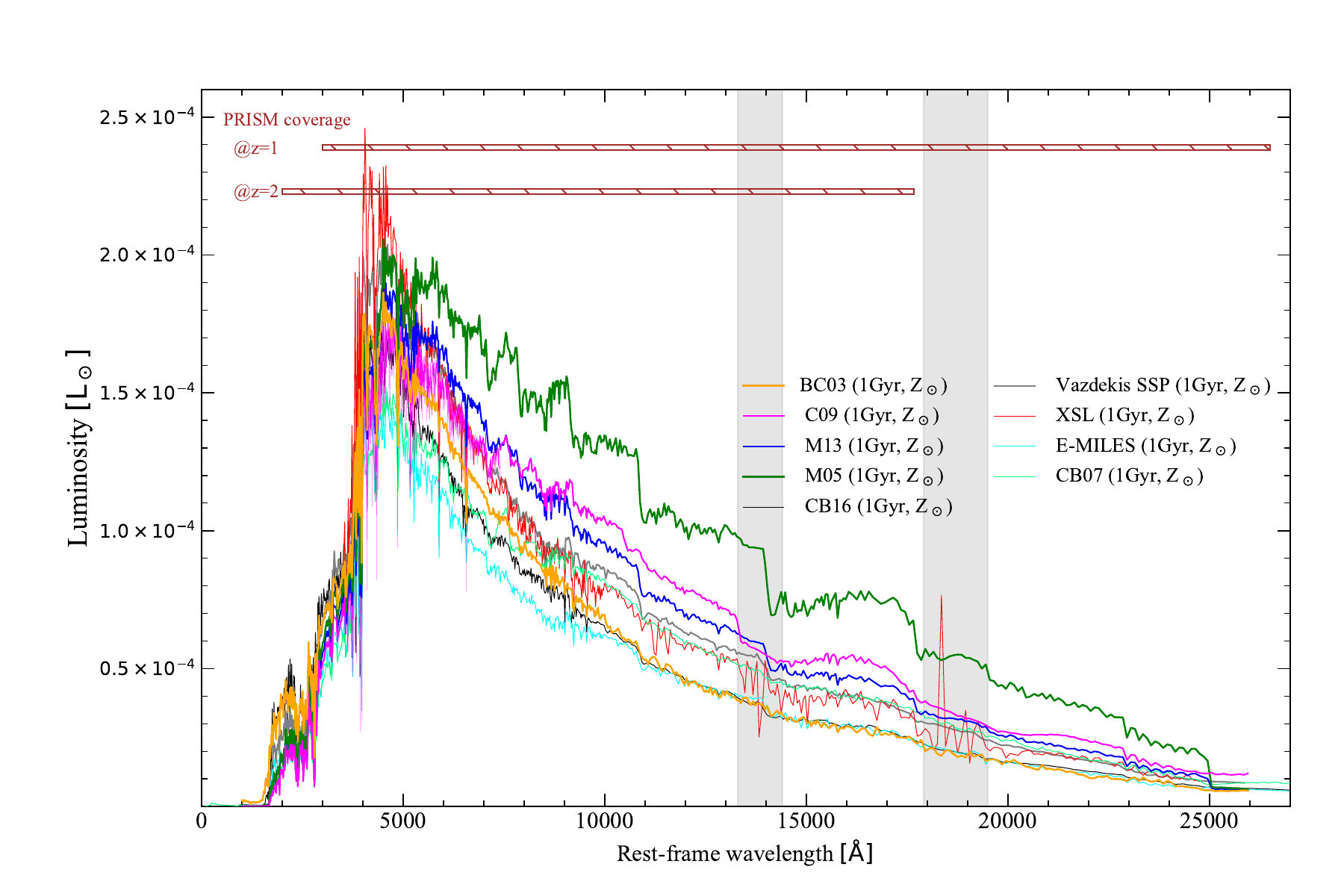}
\end{center}
\vspace{-0.3truecm}  
\caption{\small \textbf{The SEDs of different SSP models of 1 $\rm M_\odot$, $\sim$1 Gyr and solar metallicity.} M05/M13 models include a larger contribution from cool C- and O-rich stars compared to the BC03/C09 models. The CB07 (CB16) model is the 2007 (2016) version of the BC03 model. The Vazdekis SSP is constructed via isochrone synthesis, and contains all phases but excludes the TP-AGB phase. Additional SSPs, from CB16, Vazdekis, XSL, E-MILES, and CB07 are smoothed to match the sampling of M13 model for better visualization. Two noisy telluric regions in the XSL model are shown in grey-shaded regions. The growing NIR output can be viewed as a gauge of decreasing mass-to-light (M/L) ratios. The rest-frame PRISM coverages at $z=$1 and $z=$2, respectively, are depicted at the top.}\label{SI_fig03}
\vspace{-12pt}
\end{figure*}

\begin{small}
\begin{table*}
\setlength{\tabcolsep}{2pt}
\centering
\caption{\textcolor{black}{List of proposed NIR features associations in D36123.}}\label{SI_tab1}
\begin{tabular}{c|c|c|c|c|c} 
\hline 
        &    &\multicolumn{2}{c}{This work} & \multicolumn{2}{|c}{Literature} \\
\cline{3-6}
Type   & No.& $\Delta \lambda_{\rm rest}$ [\AA] & specific name & $\Delta \lambda_{\rm rest}$ [\AA] & index name\\
\hline
&1  & 5395-5555 & CN0.55 &--&--\\
&2  & 5900-6080 & CN0.6  &--&--\\
&3  & 6921-7060 & CN0.7 &--&--\\
edge &4  & 7740-7890  & CN0.78 &--&-- \\
(C-rich) &5  & 9110-9290 & CN0.92&--&-- \\
&6  & 10830-11040 & CN1.1 & 10780-11120\cite{Riffel+09,Riffel+15,Riffel+19} & CN11\\
&7 & 13755-13885  & CN1.38 & --&--\\ 
&8 & 17660-17745  & CN1.77 & --&--\\
\hline
 & 9 & 5045-5285  & MgI+MgII+TiO$\alpha$($\Delta \nu=$0)+C$_2$? &5069-5197\cite{Worthey+1994,Riffel+19}& Mg1/Mg2/Mgb \\
 & 10 &5560-5730  & TiO$\beta$($\Delta \nu=$0)+C$_2$ &--&-- \\
 & 11 & 5900-5990 & TiO${\gamma'}$($\Delta \nu=$1) & 5937-5994\cite{Worthey+1994,Riffel+19}&TiO1\\
 & 12 & 6170-6355 & TiO${\gamma'}$($\Delta \nu=$0)+C$_2$ &6191-6272\cite{Worthey+1994,Riffel+19}&TiO2 \\
absorption & 13 & 7114-7310 & TiO$\gamma$($\Delta \nu=$0) & --&-- \\
(O-rich) & 14 &7790-7985 & VO B-X ($\Delta \nu=$0) & --&--\\
& 15 & 8455-8645 & TiO$\epsilon$($\Delta \nu=$0)+VO B-X($\Delta \nu=$-1)+CaII & 8476-8700\cite{Riffel+19} & CaT1/CaT2/CaT3\\
& 16 & 8830-9015 & TiO$\delta$($\Delta \nu=$0) & --&-- \\
& 17 & 9465-9655 & TiO$\epsilon$($\Delta \nu=$-1)+ZrO? & 9200-9500\cite{Riffel+15,Riffel+19}& CN/TiO/ZrO\\
& 18 & 10458-10675 & VO A-X ($\Delta \nu=$0) &10470-10650\cite{Riffel+15,Riffel+19} & VO\\
\hline
\end{tabular}
\begin{tablenotes}
\footnotesize 
\vspace{5pt}
\item NIR features for which we propose identification in the rest-frame spectrum of object D36123, include  (C-rich) spectral edges and (O-rich) absorptions. These rest-frame wavelength range of specific features are approximate owing to the fairly low resolution of the spectrum, and also because the broadness of the features make it harder to clearly define a center. Some specific features defined by previous studies\cite{Worthey+1994,Riffel+09,Riffel+15,Riffel+19} are listed in the right two columns for comparison. 
\end{tablenotes}
\end{table*}
\end{small}

\begin{small}
\begin{table*}
\centering
\caption{Parametric grid for the D36123 spectral fit}\label{SI_tab2}
 \begin{tabular}{lcllr} 
 \hline
\multicolumn{2}{l}{Parameters [units]} & Min & Max & Step \\
 \hline\hline
\multicolumn{2}{l}{$z$}&  1.06   & 1.1    & 0.001	 \\
 \hline
            &  & 0.1 & 3.0 & 0.1 \\
 \multicolumn{2}{l}{$t^{\dagger}$ [Gyr]}& 3.5 & 5.0 & 0.5 \\
            &  & 5.0 & 10.0 & 1.0 \\
            &  & 10.0 & 20.0 & 5.0 \\
 \hline
    &  & \multicolumn{3}{l}{0.001, 0.004, 0.009, 0.05} \\
\multicolumn{2}{l}{$\tau$ [Gyr]}& 0.1 & 1.0 & 0.1 \\
    &   & 1.0 & 3.0 & 1.0 \\
 \hline
\multicolumn{2}{l}{Av [mag]}  & 0 & 2.1 & 0.1 (Calzetti) \\
 \hline  
  &  & 0.3 & 2.5 & 0.1 (BC03)\\
  & {\bf Main} & 0.3 & 2.0 & 0.1 (C09) \\
  &  & 0.3 & 2.2 & 0.1 (M13) \\
\multicolumn{2}{l}{ $Z/Z_\odot$} & 0.5 & 2.0 & 0.1 (M05)\\
\cline{3-5}
     &      & 0.3 & 1.6 & 0.1 (XSL)\\
     &\textcolor{black}{Additional} & 0.3 & 2.5 & 0.1 (E-MILES)\\
     &      & 0.3 & 2.5 & 0.1 (CB07)\\
\hline
 \end{tabular}
\begin{tablenotes}
\footnotesize 
\vspace{5pt}
\item[]{$^{\dagger}$ indicates the time since the onset of star formation. The solar metallicity $Z_\odot=$0.02 is adopted.} 
\end{tablenotes}
\end{table*}
\end{small}

\vspace{3pt}
\noindent\rule{\linewidth}{0.4pt}
\vspace{3pt}

\begin{addendum}
\item[Data Availability]
The JWST NIRSpec data are available from the Mikulski Archive for Space Telescope (MAST;  http://archive.stsci.edu), under program IDs 1345 and 2750. The CEERS JWST imaging data are available from MAST under program ID 1345. Reduced NIRCam data products from the CEERS team are available at https://ceers.github.io. The HST imaging data are available from the CANDELS survey at https://archive.stsci.edu/hlsp/candels, and the NIRCam/Ks imaging data are published in the 3D-HST survey at https://archive.stsci.edu/prepds/3d-hst/. 

\item[Code Availability]
The JWST NIRSpec data were reduced using the JWST Pipeline (version 1.8.5, reference mapping 1041 and 1029; https://github.com.spacetelescope/jwst). The MSAFIT software is available at https://github.com/annadeg/jwst-msafit. The {\tt galfit} software is published at https://users.obs.carnegiescience.edu/peng/work/galfit/galfit.html. 

\item[Acknowledgements]
We thank L. Origlia and B. Holwerda for discussions. M.D., P.A.H., S.L.F., J.S.K. and C.P. acknowledge support from NASA through the Early Release Science Program of the Space Telescope Science Institute (Award JWST-ERS-1345) and the JWST-GO-2750 award. C.G.G. acknowledges support from the French National Centre for Space Studies. C.D.E. acknowledges funding from the State Research Agency of the Ministry for Science and Innovation (Spain) and the NextGenerationEU/PRTR (Recovery, Transformation, and Resilience Plan, European Union) through the Juan de la Cierva-Formación programme (Grant No. FJC2021-047307-I). S.L. acknowledges support from the China Scholarship Council. This work is supported by the National Natural Science Foundation of China (Grant Nos. 12192222, 12192220 and 12121003). This work is based on observations with the NASA/ESA/CSA JWST obtained from the Mikulski Archive for Space Telescopes at the Space Telescope Science Institute, which is operated by the Association of Universities for Research in Astronomy, Incorporated, under NASA contract NAS5-03127.

\item[Author Contributions Statement ]
SL and ED conceived this project, and selected and identified galaxies. SL led the analysis and together with ED, CM and MD wrote the manuscript. PAH and MD led the original observations and reduced the NIRSpec spectra. RG wrote the fitting code and with CDE helped with the spectral fitting procedures. All authors aided in the analysis and interpretation and contributed to the final manuscript.

\item[Competing Interests Statement]
The authors declare no competing interests.

\item[Additional information]
Correspondence and requests for materials should be addressed to Shiying Lu (email: ShiyingLu@smail.nju.edu.cn), Emanuele Daddi (emanuele.daddi@cea.fr) and Claudia Maraston(claudia.maraston@port.ac.uk).

\end{addendum}

\vspace{3pt}
\noindent\rule{\linewidth}{0.4pt}
\vspace{3pt}

\newcommand{\noop}[1]{}

\end{document}